\documentclass[aps,prd,twocolumn,groupedaddress,showpacs,floatfix]{revtex4}
\usepackage[dvipdfm]{graphicx}
\usepackage{epsf}
\usepackage{amsmath,amssymb}

\newcommand{\beq}{\begin{equation}}
\newcommand{\eeq}{\end{equation}}
\newcommand{\beqn}{\begin{eqnarray}}
\newcommand{\eeqn}{\end{eqnarray}}
\newcommand{\pa}{\partial}

\newcommand{\varep}{\varepsilon}

\newcommand{\brr}{{\mbox{\boldmath$r$}}}

\begin{document}
\title{Quasiequilibrium states of black hole-neutron star binaries in the 
moving-puncture framework}
\author{Koutarou Kyutoku, Masaru Shibata}
\affiliation{Yukawa Institute for Theoretical Physics, Kyoto University,
Kyoto, 606-8502, Japan}
\author{Keisuke Taniguchi}
\affiliation{Department of Physics, University of Wisconsin-Milwaukee,
P.O. Box 413, Milwaukee, Wisconsin 53201, USA}
\date{\today}

\begin{abstract}
 General relativistic quasiequilibrium states of black hole-neutron
 star binaries are computed in the moving-puncture framework. We
 propose three conditions for determining the quasiequilibrium
 states and compare the numerical results with those obtained in the
 excision framework.  We find that the results obtained in the
 moving-puncture framework agree with those in the excision framework
 and with those in the third post-Newtonian approximation for the cases
 that (i) the mass ratio of the binary is close to unity irrespective of
 the orbital separation, and (ii) the orbital separation is large enough
 ($m_0\Omega \alt 0.02$, where $m_0$ and $\Omega$ are the total mass and
 the orbital angular velocity, respectively) irrespective of the mass
 ratio. For $m_0 \Omega \agt 0.03$, both of the results in the
 moving-puncture and excision frameworks deviate, more or less, from
 those in the third post-Newtonian approximation. Thus the numerical
 results do not provide a quasicircular state, rather they seem to have
 a non-negligible eccentricity of order $0.01$--0.1. We show by
 numerical simulation that a method in the moving-puncture framework 
 can provide approximately quasicircular states in which the
 eccentricity is by a factor of $\sim 2$ smaller than those in 
 quasiequilibrium given by other approaches.
\end{abstract}

\maketitle

\section{introduction} 

Ground-based laser interferometric gravitational-wave detectors such
as LIGO \cite{LIGO}, VIRGO \cite{VIRGO}, GEO600 \cite{GEO}, and TAMA300
\cite{TAMA} are now in operation, and advanced detectors such as
advanced LIGO will be in operation in the next decade and are expected
to detect gravitational waves. Among many other sources, coalescing
binary compact objects such as neutron star-neutron star and black
hole-neutron star (hereafter NS-NS and BH-NS, respectively) binaries are
the most promising sources \cite{voss,kalogera,belczynski}. This has
been motivating theoretical studies of gravitational waves from the
inspiral and merger of the binary compact objects.

In the past two decades, a number of short-hard $\gamma$-ray bursts
has been observed by the $\gamma$-ray and x-ray satellites
\cite{nakar,leeramirez}.  However, the progenitors of these bursts are still
highly uncertain. One of the plausible models of the central engine is
the merger of NS-NS and/or BH-NS binaries
\cite{npp1992,fryer1999}. This scenario is based on
the idea that a system consisting of a rotating BH and hot, massive
accretion-disk is formed as a consequence of the merger, and
subsequently, they emit huge amount of $\gamma$-rays and x-rays in a
short time scale.  To theoretically explore this possibility (more
specifically, to clarify the formation process of the BH-disk system),
general relativistic study for the merger of NS-NS and BH-NS binaries is
probably the unique approach. This issue has been also motivating
numerical studies for the merger of NS-NS and BH-NS binaries.

In the past decade, substantial effort has been paid in the community
of numerical relativity for clarifying the inspiral and merger
processes of binary compact objects. In particular, a wide variety of
simulations have been performed for the merger of NS-NS binaries
\cite{NSNS1,NSNS2,NSNS3,NSNS4,NSNS5,NSNS6,NSNS7,NSNS8,NSNS9,NSNS10}
and black hole-black hole (hereafter BH-BH) binaries
\cite{BHBH1,BHBH2,movingpunc,BHBH4,BHBH5,BHBH6,BHBH7,BHBH8,BHBH9,BHBH10,BHBH11,BHBH12,BHBH13,BHBH14,BHBH15,BHBH16,BHBH17,BHBH18,BHBH19,BHBH20,BHBH21,BHBH22,BHBH23,BHBH24,BHBH25,BHBH26,BHBH27,BHBH28,BHBH29}.
In the past two years, a method for computing quasiequilibrium states
of BH-NS binaries has been developed \cite{grandclement2006,tbfs2006,KT_ex,KT_ex2,foucart}, and
also, numerical simulations for the merger of BH-NS binaries have been
done \cite{BHNSpunc,BHNSpunc2,BHNS,BHNS2,BHNS3,SACRA}. However, these
researches are still in early stage; e.g., most of the simulations
have been performed only for a short time scale, and for such
simulations, the inspiral and subsequent merger phases are not likely to
be computed very accurately (but see \cite{BHNS4,BHNS08}).  Also, it is
not clear whether the computed quasiequilibrium states are really
quasicircular; i.e., it is not clear whether the eccentricity is
sufficiently small. Numerical simulations are performed employing the
quasiequilibrium states as the initial condition. If the
quasiequilibrium state is not in a quasicircular orbit, the results of
the numerical simulation are not realistic. The purpose of this paper is
(i) to present an improved study for the quasiequilibrium states and
(ii) to compare several quasiequilibrium states obtained so far. In an
accompany paper \cite{BHNS08}, we also present the latest accurate
numerical results for the inspiral and merger of BH-NS binaries. 

To compute a quasiequilibrium state of BH-NS binaries, we have to employ
an appropriate method by which the singular behavior of the BH is
avoided. In most of the previous studies
\cite{grandclement2006,tbfs2006,KT_ex,KT_ex2,foucart}, the
quasiequilibrium state is computed in the so-called excision framework,
in which a spherical region inside an apparent horizon is excised and
basic equations are solved imposing plausible boundary conditions at the
excised two-sphere. In this paper, we employ the so-called
moving-puncture approach (see Sec.~\ref{sec:field}), which is an
alternative of the excision approach \cite{BHNSpunc,BHNSpunc2}.  In this
approach, we do not have to excise any region around the BH horizon nor
impose boundary conditions around the BH, as shown in
Ref.~\cite{punc}. Indeed, this has been proven to be quite useful for
computing a quasiequilibrium state \cite{baumgarte}  and also for
simulating binary BHs (e.g., \cite{movingpunc,BHBH7}).

Another possible merit in the moving-puncture approach is that there
is a flexibility for computing a quasiequilibrium state. In the
excision method, one imposes the boundary conditions at the excised
two-sphere and at spatial infinity. The boundary conditions at the
excised surface are usually determined by requiring that the two-sphere
should be the Killing horizon at least approximately.  As a result, a
quasiequilibrium state is completely determined with no ambiguity,
although it is not clear whether the obtained quasiequilibrium is really
a quasicircular state \cite{BERTI}.  By contrast, any boundary condition
does not have to be imposed around the BH in the moving-puncture
approach. Because of this, we have a remaining degree of freedom for
defining a quasiequilibrium state. Specifically, we do not have any
natural, physical condition for determining the center of mass of the
system in this method. However, this degree of freedom can be used to
obtain a favorable, quasiequilibrium state. As we illustrate in
Sec.~\ref{sec:numeric}, quasiequilibrium states obtained in the excision
method are not quasicircular in general: Namely, the eccentricity is not
zero (see also Ref.~\cite{BERTI}). In the moving-puncture approach, the
remaining degree of freedom can be used to reduce the eccentricity, and
it may be possible to obtain a quasiequilibrium state in which the
eccentricity is smaller than that obtained by the excision method.

This paper is organized as follows: In Sec.~\ref{sec:formulation}, we
first review the basic equations for computing quasiequilibrium states
in the moving-puncture approach. Then, we describe three methods for
determining the center of mass in the moving-puncture approach.  The
numerical methods for solving a quasiequilibrium state are described in
Sec.~\ref{sec:numeric}. We present our numerical results and compare
those with other results in Sec.~\ref{sec:results}. In
Sec.~\ref{sec:simulation}, we present results of numerical simulation of
the inspiral and merger of BH-NS binaries for a particular model: We
adopt two initial data obtained in the moving-puncture approach, and
compare the numerical results. We then demonstrate that one of the
moving-puncture approach is superior for the initial condition because
the eccentricity is smaller than those obtain in other methods. Section
\ref{sec:discussion} is devoted to a summary. Throughout this paper, we
adopt the geometrical units in which $G=c=1$, where $G$ and $c$ are the
gravitational constant and the speed of light. Latin and Greek indices
denote spatial and spacetime components, respectively. 

\section{Formulation} \label{sec:formulation}

In this section, we first review the basic equations for computing a
BH-NS binary in quasiequilibrium, and then describe the quantities
used in the analysis and the methods for defining the quasiequilibrium
state in the moving-puncture approach.

\subsection{Field equations in the moving-puncture framework}
\label{sec:field}

To derive a quasiequilibrium state of BH-NS binaries as a solution of
the initial value problem of general relativity, we employ a mixture of
the conformal thin-sandwich decomposition \cite{TS} and conformal
transverse-traceless decomposition of the Einstein equations. Following
the previous works, we assume that the trace part of the extrinsic
curvature ($K=$tr$(K_{ij})$) is zero and the three-metric
($\gamma_{ij}$) is conformally flat \cite{rev_initial}:
$\gamma_{ij}=\psi^4 f_{ij}$, where $\psi$ is the conformal factor and
$f_{ij}$ is the flat metric. 

We define a tracefree, weighted extrinsic curvature as
\begin{equation}
\hat{A}^{ij} =\psi^{10} K^{ij}.
\end{equation}
Because we assume that the three-metric is conformally flat, this
quantity is written by 
\begin{equation}
\hat{A}^{ij} = \frac{\psi^6}{2\alpha}\left( \hat{\nabla}^i \beta^j +
				      \hat{\nabla}^j \beta^i -
				      \frac{2}{3} f^{ij}\hat{\nabla}_k\beta^k
				      \right),\label{hataij0}
\end{equation}
where $\alpha$ is the lapse function, $\beta^k$ the shift vector, and 
$\hat \nabla_i$ the covariant derivative with respect to $f_{ij}$. Note
that adding a rotational shift vector,
\begin{equation}
 \beta^i_{\rm rot} = ({\bf \Omega} \times {\bf R})^i ,
\end{equation}
does not change $\hat{A}^{ij}$ in the conformal flatness
formalism. Here, $\Omega$ is the angular velocity vector of the binary
and $R$ is the coordinate vector from the center of mass of the
binary. Thus, computation may be performed in any rotational frame using
the same equations, by simply changing the boundary conditions.

In the present formalism, the basic equations are derived from the
Hamiltonian constraint, momentum constraint, and the maximal slicing
condition ($\pa_t K=0$) \cite{wilsonmathews1995} as
\begin{eqnarray}
&& \Delta \psi = -2\pi\psi^5\rho_H -
 \frac{1}{8}\psi^{-7}\hat{A}_{ij}\hat{A}^{ij}, \label{eq:psi} \\
&&\Delta \beta^i + \frac{1}{3}\hat{\nabla}^i\hat{\nabla}_j\beta^j
 = 16\pi\alpha\psi^4 j^i 
+ 2\hat{A}^{ij}\hat{\nabla}_j(\alpha\psi^{-6}), \label{eq:beta} \\
&&\Delta\Phi = 2\pi\Phi\psi^4(\rho_H + 2S_k^{~k}) +
 \frac{7}{8}\Phi\psi^{-8}\hat{A}_{ij}\hat{A}^{ij},
 \label{eq:alpha}
\end{eqnarray}
where $\Delta = f^{ij}\hat{\nabla}_i\hat{\nabla}_j$, $\Phi \equiv
\alpha\psi$, and 
\begin{eqnarray}
&&\rho_H = T^{\mu\nu} n_\mu n_\nu , \\
&&j^i = - T^{\mu\nu} n_\mu \gamma^i_\nu , \\
&&S_{ij} = T^{\mu\nu}\gamma_{i\mu}\gamma_{j\mu}.
\end{eqnarray}
Here, $n^{\mu}$ is the timelike hypersurface normal,
$\hat A_{ij}=f_{ik}f_{jl} \hat A^{kl}$, and $T^{\mu \nu}$ the
stress-energy tensor.

For computing a quasiequilibrium state of a system containing BHs, we
have to appropriately treat singular behaviors of the BHs because
divergent quantities cannot be handled in numerical computation. Most of
the previous works for computing quasiequilibrium states of BH-NS
binaries \cite{grandclement2006,tbfs2006,KT_ex,KT_ex2,foucart} have been
done with an ``excision approach,'' i.e., excising the region inside
two-sphere of apparent horizon from the computational domain with
appropriate boundary conditions at the excision surface. On the other
hand, a ``puncture'' method \cite{punc} was proposed by Brandt and
Br{\"u}gmann to describe multiple BHs with arbitrary linear momenta and
spin angular momenta, and a ``moving-puncture approach''
\cite{movingpunc,BHBH7} was revealed to be quite useful in dynamical
simulations. In this paper, we employ the moving-puncture approach,
which is developed by Shibata and Ury{\=u} \cite{BHNSpunc,BHNSpunc2} for
the case of BH-NS binaries. In the puncture or moving-puncture framework
we decompose the metric quantities into a singular part, which is
written analytically and denotes contribution from a BH, and a regular
part, which is obtained by numerically solving the basic
equations. Assuming that the puncture is located at $\brr_{\rm
P}=x^k_{\rm P}$, we set $\psi$ and $\Phi$ as
\begin{eqnarray}
&&\psi = 1 + \frac{M_{\rm P}}{2r_{\rm BH}} + \phi, \\
&&\Phi = 1 - \frac{M_{\rm \Phi}}{r_{\rm BH}} + \eta,
\end{eqnarray}
where $M_{\rm P}$ and $M_{\rm \Phi}$ are positive constants of mass
dimension, and $r_{\rm BH} = |x^k - x^k_{\rm P}|$ is a coordinate
distance from the puncture. $M_{\rm P}$ is an arbitrarily chosen
parameter called the puncture mass, whereas $M_{\rm \Phi}$ is
determined by the condition that Arnowitt-Deser-Misner (ADM) mass
($M_0$), and Komar mass agree (this condition should hold when the
spacetime is stationary and asymptotically flat
\cite{virial_beig,virial_ash}), i.e., 
\begin{equation}
\oint_{r\rightarrow \infty}\partial_i \Phi dS^i = - \oint_{r\rightarrow
 \infty}\partial_i \psi dS^i = 2\pi M_0. \label{eq:condition_C}
\end{equation}
Also, we decompose $\hat A_{ij}$ into singular and regular parts as 
\begin{equation}
\hat{A}_{ij} = \hat \nabla_i W_j + \hat \nabla_j W_i -
 \frac{2}{3}f_{ij}\hat \nabla_k W^k + K^{\rm P}_{ij}, \label{hataij}
\end{equation}
where $K^{\rm P}_{ij}$ is the singular part, which denotes a weighted
extrinsic curvature associated with the linear momentum of the BH
written by 
\begin{equation}
K^{\rm P}_{ij} = \frac{3}{2r^2_{\rm BH}}\Big[l_i P^{\rm BH}_j + l_j
 P^{\rm BH}_i -(f_{ij} - l_i l_j)l^k P^{\rm BH}_k \Big],
\end{equation}
and $l^k = x^k_{\rm BH}/r_{\rm BH}$. $W_i$ denotes an auxiliary
three-dimensional function and $W^i =f^{ij} W_j$. Because the total
linear momentum of the system should vanish, the linear momentum of the
BH, $P_i^{\rm BH}$, is related to that of the companion NS as
\begin{equation}
P^{\rm BH}_i = - \int j_i \psi^6 dV, \label{eq:mom_punc}
\end{equation} 
where the right-hand side denotes the (minus) linear momentum of the
NS.

Field equations that we have to solve are summarized as follows: 
\begin{eqnarray}
&&\Delta\phi = -2\pi\psi^5\rho_H -
 \frac{1}{8}\psi^{-7}\hat{A}_{ij}\hat{A}^{ij}, \label{eq:phi_fin} \\
&&\Delta\beta^i+\frac{1}{3}\hat \nabla^i \hat \nabla_j \beta^j =
 16\pi\Phi\psi^3j^i + 2\hat{A}^{ij}\hat \nabla_j(\Phi\psi^{-7}),
 \label{eq:beta_fin} \\
&&\Delta\eta = 2\pi\Phi\psi^4(\rho_H + 2S_k^{~k}) +
 \frac{7}{8}\Phi\psi^{-8} \hat{A}_{ij}\hat{A}^{ij}, \label{eq:eta_fin}
 \\ 
&&\Delta W_i + \frac{1}{3}\hat \nabla_i \hat \nabla_j W^j = 8\pi
 \psi^{6} j_i.
\label{eq:W_fin}
\end{eqnarray}
We note that $\hat A_{ij}$ is obtained by Eq.~(\ref{hataij}), not by
Eq.~(\ref{hataij0}), because $\hat A^{ij}$ is not straightforwardly
defined for $\alpha=0$ when we adopt Eq.~(\ref{hataij0}). In this
approach, the elliptic equation for $\beta^i$ has to be solved because
we need $\beta^i$ in solving hydrostatic equations (see
Sec.~\ref{sec:hydro}).

All the basic equations are elliptic type, and hence, we have to
impose appropriate boundary conditions at spatial infinity. 
Because of the asymptotic flatness, 
the boundary conditions at spatial infinity $r \rightarrow \infty$ 
are written as
\begin{equation}
 \phi , \beta^i , \eta , W^i \rightarrow 0,
\end{equation}
where we assume that the equations are solved in the inertial frame. We
note that outer boundaries are located at spatial infinity  in our
numerical computation (cf. Sec.~\ref{sec:numeric}).  Thus, the above
condition is exactly imposed. 

In contrast to the case that the excision approach is adopted, we do
not have to impose the inner boundary conditions in the moving-puncture
approach. This could be a drawback in this approach, because we cannot
impose physical boundary conditions (e.g., Killing horizon boundary
condition) for the BH. However, this could be also a merit, because we
have a flexibility for adjusting a quasiequilibrium state to a desired
state by using this degree of freedom. In Sec.~\ref{sec:free}, we
discuss this point in more detail.

\subsection{Hydrostatic equations} \label{sec:hydro}

Assuming that the NS is composed of an ideal fluid, we write the
stress-energy tensor as
\begin{equation}
T_{\mu\nu} = (\rho + \rho \varepsilon + p)u_\mu u_\nu + p g_{\mu\nu},
\end{equation}
where $\rho$ is the baryon rest-mass density, $\varepsilon$ the
specific internal energy, $p$ the pressure, and $u_\mu$ the fluid
four-velocity.  We employ a polytropic equation of state
\begin{equation}
p = \kappa \rho^\Gamma,
\end{equation}
where $\kappa$ is the polytropic constant and $\Gamma$ the adiabatic
index. In this paper, we set $\Gamma = 2$ following previous works 
\cite{grandclement2006,tbfs2006,KT_ex,KT_ex2,foucart}. Using the
first law of thermodynamics, $\varepsilon$ is obtained as
$p/[(\Gamma-1)\rho]$, and  the specific enthalpy, $h=1+\varep+p/\rho$,
is given by
\begin{equation}
h = 1 + \kappa \frac{\Gamma}{\Gamma - 1}\rho^{\Gamma -1}.
\end{equation}
Thus, all the thermodynamic quantities are written in terms of $\rho$. 

In the polytropic equation of state, all the dimensional quantities
enter the problem only through the polytropic constant, and thus, are
rescaled into a dimensionless form by normalizing in terms of the
polytropic length scale,
\begin{equation}
R_{\rm poly} \equiv \kappa^{1/(2\Gamma -2)}.
\end{equation}
In this paper, we present all the quantities in the dimensionless form
following Refs.~\cite{grandclement2006,tbfs2006,KT_ex,KT_ex2,foucart}. 

BH-NS binaries are never in a true equilibrium due to the emission of 
gravitational waves. However, when the orbital separation $d$ is large 
enough, emission time scale of gravitational waves $t_{\rm GW}$ is much
longer than the orbital period $t_{\rm orb}$ as
\begin{equation}
 \frac{t_{\rm GW}}{t_{\rm orb}} \approx 
 1.1\left(\frac{d}{6m_0}\right)^{5/2}\left(\frac{m_0}{4\mu}\right),
\end{equation}
where $m_0$ and $\mu$ denote the total mass and reduced mass of the
binary.  Thus, except for the final inspiral phase, say $d < 10m_0$, the
effect of gravitational radiation reaction may be safely neglected, and
the binary can be regarded to be approximately in an equilibrium
state. Because the binaries in a close orbit should have a circular
orbit \cite{petersmathews1963,peters1964}, the fluid configuration
should be in hydrostatic equilibrium in the corotating frame of the
binary system. In addition, it is believed that the matter in most of
the NS has the approximately irrotational velocity field for the
realistic binary configurations because the viscous time scale for the
angular momentum transport inside the NS is much longer than the
gravitational radiation reaction time scale \cite{irrot,irrot2}. (We
note that the actual NSs are known to have nonzero spin angular momenta
and not exactly in the irrotational states. However, we can still
approximate astrophysical NSs well with the irrotational velocity fields
because their typical rotational period is 100ms-1s and are much longer
than their typical orbital period just before mergers, $\sim 2-3$
ms, and also much longer than the dynamical time scale of the NSs,
$\lesssim 1$ ms.)

The equations of relativistic hydrostatic equilibrium with the
irrotational velocity field is derived independently in
Refs.~\cite{ir_S,ir_T,ir_B}, which are summarized in
Ref.~\cite{irrot_hydro}. In this formulation, one assumes the presence
of a helical Killing vector
\begin{equation}
\xi^\mu = (\partial_t)^\mu + \Omega (\partial_\phi)^\mu.
\end{equation}

For the irrotational velocity field, the relativistic vorticity is zero
as
\begin{equation}
\omega_{\mu\nu}=\nabla_\mu(h u_\nu) - \nabla_\nu(h u_\mu) = 0. 
\label{eq:vorticity}
\end{equation}
Using Eq.~(\ref{eq:vorticity}) and the helical symmetric relation for
the specific momentum $\pounds_\xi (h u^\mu) = 0$, we obtain the first
integral of the relativistic Euler equation as
\begin{equation}
h \xi_\mu u^\mu = -C(={\rm const}). \label{eq:firstinteg}
\end{equation}
To rewrite this equation into a more specific form, we decompose the
four-velocity in the form
\beqn
u^{\mu}=u^t (\xi^{\mu} + V^{\mu}),
\eeqn
where $V^{\mu}$ is a three-velocity (i.e., $n_{\mu}V^{\mu}=0$) 
and denotes the velocity field in the comoving frame of the 
binary system. Then, $\xi^{\mu}$ is written as 
$\xi^{\mu}=u^{\mu}/u^t-V^{\mu}$. Substituting this 
equation into Eq.~(\ref{eq:firstinteg}), we obtain \cite{ir_S}
\begin{equation}
{h \over u^t} + \tilde u_i V^i  = C, \label{eq:rel_euler}
\end{equation}
where $\tilde u_i=h \gamma_i^{~\mu} u_{\mu}$ denotes the three 
specific momentum. 

The condition of irrotation, Eq.~(\ref{eq:vorticity}), implies that
$\tilde u_i$ is written by the gradient of a velocity potential field
$\Psi$ such that
\begin{equation}
\tilde u_i = D_i \Psi,
\end{equation}
where $D_i$ is the covariant derivative with respect to $\gamma_{ij}$. 
Then, $V^i$ is written by the velocity potential as
\beqn
V^i=-\xi^i-\beta^i + {1 \over h u^t}D^i \Psi, \label{eq34}
\eeqn
and also
\beqn
u^t={1 \over \alpha} \Bigl[1+h^{-2} D^k \Psi D_k \Psi \Bigr]^{1/2}.
\eeqn
Thus, the first integral of the Euler equation is written 
by $h$, $\Psi$, and geometrical quantities. 

The equation for the velocity potential is derived from the 
continuity equation, which can be written in the presence of 
the helical Killing vector as \cite{ir_S}
\beqn
D_i (\rho \alpha u^t V^i) = 0. \label{eq36}
\eeqn
Substituting Eq.~(\ref{eq34}) into Eq.~(\ref{eq36}), an elliptic-type
equation for $\Psi$ is derived to give
\begin{equation}
D_i [\rho \alpha \{ h^{-1} D^i \Psi - u^t (\xi^i + \beta^i) \} ]=0.
\label{velpoteq}
\end{equation}
This equation is also written by $h$, $\Psi$, and geometrical
quantities. Thus, from the first integral of the Euler equation and 
Eq.~(\ref{velpoteq}), $h$ and $\Psi$ are computed, and subsequently, 
$\rho$, $\varep$, and $p$ are determined from the equation of state. 

\subsection{Global quantities} \label{sec:global}

A quasiequilibrium state is characterized by the mass and spin of the
BH, the mass and radius of the NS, and the orbital angular velocity
$\Omega$. A quasiequilibrium sequence should be a sequence as a function
of $\Omega$ with constant values of the BH mass, the BH spin, and the
baryon rest mass of the NS, and with a fixed equation of state for the
NS. In this paper, we assume that the BH spin is zero, and the BH mass
is defined by the irreducible mass as
\begin{equation}
M^{{\rm BH}}_{{\rm irr}} = \sqrt{\frac{A_{{\rm EH}}}{16\pi}},
\end{equation}
where $A_{{\rm EH}}$ is the proper area of the event horizon. In 
practice, we approximate this area with that of the apparent
horizon, which is computed from an integral on the apparent-horizon 
surface
\begin{equation}
A_{{\rm AH}} = \int_{{\rm AH}} \psi^4 dS,
\end{equation}
where we use that the three-metric is conformally flat.  In the
moving-puncture approach, in contrast to the excision approach in which
the two-sphere of the apparent horizon is readily known to be the
excision surface, the apparent horizon has to be determined by a
numerical computation. For finding the apparent horizon, we use the
algorithm developed by Lin and Novak (see Ref.~\cite{LoreneAH} for
details).

The baryon rest mass of the NS is defined by 
\begin{equation}
M^{{\rm NS}}_{{\rm B}} = \int \rho u^t \sqrt{-g} dV,
\end{equation}
where $g$ is the determinant of the spacetime metric $g_{\mu\nu}$. 
In this paper, we always present the mass in polytropic units as 
\begin{equation}
\bar M^{\rm NS}_{\rm B} \equiv
 \frac{M^{{\rm NS}}_{{\rm B}}}{M_{\rm poly}},
\end{equation}
whereby $\bar M^{\rm NS}_{\rm B}$ is the baryon rest mass for the
polytropic constant $\kappa = 1$.

In the following, we often compare numerical results with those
derived by the third post-Newtonian (3PN) approximation for two points
masses \cite{3PN,3PN2}.  In such a case, we have to define the total
mass $m_0$ and the reduced mass $\mu$ of the system (i.e., we have
to define each mass of the binary component). For the BH mass, we use
the irreducible mass. For the NS mass, we use the ADM mass of an
isolated NS $M^{\rm NS}_{\rm ADM}$ with the same baryon rest
mass. Thus,
\beqn
&&m_0=M_{\rm irr}^{\rm BH} + M^{\rm NS}_{\rm ADM}, \\
&&\mu={M_{\rm irr}^{\rm BH} M^{\rm NS}_{\rm ADM} \over m_0}.
\eeqn
Note that for a nonspinning BH, $M_{\rm irr}^{\rm BH}$ is equal to
the ADM mass of the BH in isolation.

The ADM mass of the whole system $M_0$ is defined by 
\begin{equation}
M_0 = -\frac{1}{2\pi}\oint_{r \rightarrow \infty} \partial_i \psi dS^i,
\end{equation}
and the Komar mass is
\begin{eqnarray}
M_{\rm Komar}&=&\frac{1}{4\pi}\oint_{r \rightarrow \infty} \partial_i
\alpha dS^i \\ &=&\frac{1}{4\pi}\oint_{r\rightarrow\infty} (\partial_i \Phi -
\partial_i \psi) dS^i.  
\end{eqnarray}
Note that equating these two masses results 
in Eq.~(\ref{eq:condition_C}).
We also define the binding energy of the binary as
\begin{equation}
E_b = M_0 - m_0. 
\end{equation}

The ADM linear momentum of the system is
\begin{equation}
P_i=\frac{1}{8\pi}\oint_{r\rightarrow\infty} K_{ij} dS^j,
\end{equation}
where we assume the maximal slicing $K=0$. This is set to be zero in
the present work. The angular momentum of the system around the center
of mass of the binary may be defined by
\begin{equation}
J_i=\frac{1}{16\pi}\epsilon_{ijk}\oint_{r\rightarrow\infty} 
(X^j K^{kl}-X^k K^{jl})dS_l,
\end{equation}
where $X^i$ is the coordinate vector from the center of mass.

\subsection{Free parameters} \label{sec:free}

To compute a quasiequilibrium configuration for a given equation of
state, we have to fix free parameters. Each configuration is
determined when we fix (i) the baryon rest mass of the NS, $M^{\rm
NS}_{\rm B}$, (ii) the mass ratio, $Q$, or equivalently, the irreducible
mass of the BH, $M^{\rm BH}_{\rm irr}$, and (iii) the separation between
the BH and the NS, $d$. Then, the other parameters such as $M_{\rm P}$,
$M_{\rm \Phi}$, and $P^{\rm BH}_i$ are automatically determined in the
computation: The puncture mass $M_{\rm P}$ is arranged to give a desired
irreducible mass, the mass parameter $M_{\rm \Phi}$ is determined by the
condition that the ADM mass and the Komar mass agree (see
Eq.~(\ref{eq:condition_C})), and the linear momentum of the BH, $P^{\rm
BH}_i$, is determined by the condition that the total linear momentum of
the system should vanish []see Eq.~(\ref{eq:mom_punc})].

There are also free parameters associated with the configuration of
the NS; the integration constant $C$, which appears in the first
integral of the Euler Eq. (\ref{eq:rel_euler}), and the angular
velocity of the binary $\Omega$.  These are determined by fixing the
configuration of the binary and the baryon rest mass of the
NS. Specifically, we fix the rest mass of the NS to determine $C$ and
fix the location of the center of the NS to determine $\Omega$. Here,
the center of the NS is defined as the position at which the following
condition is satisfied \cite{KT_ex,irrot_hydro}:
\begin{equation}
\left. \frac{\partial \ln h}{\partial X}\right|_{(X_{\rm NS},Y_{\rm
 NS},0)}=0, \label{eq:omega}
\end{equation}
where $X_{\rm NS}$ and $Y_{\rm NS}$ are the distances in the $x$ and
$y$ directions from the center of the NS to the rotational axis,
respectively. In the actual calculation, we arrange the value of the
specific enthalpy at the center of the NS, $h_c$, instead of $C$, since
it is easier to implement. The value of $C$ is determined by $h_c$ and
the values of metric quantities at the center of the NS.

The final remaining task in the moving-puncture framework is to
determine the center of mass of the system. The issue in this
framework is that we do not have any natural, physical condition for
determining it.  (By contrast, the condition is automatically derived in
the excision framework
\cite{grandclement2006,tbfs2006,KT_ex,KT_ex2,foucart}, although it is
not clear whether the resulting quasiequilibrium is a quasicircular
state \cite{BERTI}.)  In our previous paper \cite{BHNSpunc,BHNSpunc2},
we employed a condition that the dipole part of $\psi$ at spatial
infinity is zero (hereafter we refer to this condition as ``dipole
condition''). However, we found that in this condition, the angular
momentum derived for a close orbit of $\Omega m_0 \agt 0.03$ is by $\sim
2\%$ smaller than that derived by the 3PN approximation \cite{3PN,3PN2}
for $Q=3$. Because the 3PN approximation should be an excellent
approximation of general relativity for a fairly distant orbit as
$\Omega m_0 \approx 0.03$, the obtained initial data deviates from the
true quasicircular state, and hence, the initial orbit would be
eccentric.

In the subsequent work \cite{BHNS}, we adopted a condition that the
azimuthal component of the shift vector $\beta^{\varphi}$ at the
location of the puncture ($\brr=\brr_{\rm P}$) is equal to $-\Omega$;
i.e., we imposed a corotating gauge condition at the location of the
puncture. In the following, we refer to this condition as the
``$\beta^{\varphi}$ condition.'' In this paper, we first present the
numerical results in the latter condition because it is a physical
condition and gives a slightly better result than the dipole condition
does.

As shown in Sec.~\ref{sec:results}, however, the angular momentum
derived for a close orbit of $\Omega m_0 \agt 0.03$ in this method is
still by $\sim 2\%$ smaller than that derived by the 3PN relation for a
large mass ratio $Q \geq 2$. The disagreement is larger for the larger
mass ratio. Such initial conditions are likely to deviate from the true
quasicircular state and hence the orbital eccentricity is large
(e.g. Sec. IV C of Ref.~\cite{SACRA} for numerical evolution of such
initial data).  This also suggests that the $\beta^{\varphi}$ condition
is not suitable for deriving realistic quasicircular states.

In this paper, we further propose a new condition in which the center
of mass is determined in a phenomenological manner: We impose the
condition that the total angular momentum of the system for a given
value of $\Omega m_0$ agrees with that derived by the 3PN
approximation. This can be achieved by appropriately choosing the
position of the center of mass.  With this method, the drawback in the
previous two methods (i.e., the angular momentum becomes smaller than
the expected value) is overcome.  We refer to this method as ``3PN-J
condition'' in the following.


\section{Numerical methods} \label{sec:numeric}

In this section, we summarize our numerical scheme for 
computing a quasiequilibrium state. 

\subsection{Methods of computation} \label{sec:method}

Our numerical code is based on the spectral methods library LORENE
\cite{Lorene}. In the spectral methods, any quantity is denoted by the
expansion into a complete set of polynomials. The feature in the
spectral methods is that the error of this expansion decreases
exponentially as the number of the used complete set of polynomials
increases, at least for continuous functions. Furthermore, irregular
functions like the rest-mass density of the NS, for which the spatial
derivative jumps at its surface and is not straightforward to expand
into a complete set of polynomials (known as ``Gibbs phenomenon''), can
be treated appropriately by employing a multidomain spectral method. We
use two sets of spherical-like computational domains, each of which is
centered on each object. The one for which the domain center is located
at the puncture is composed of one nucleus, which is sphere at the
center, several shells, and the external domain, which extends to
spatial infinity. The other has almost the same structure, except that
the outer boundary of the nucleus is deformed to fit the surface of the
NS. With this computational domain, irregular profiles of the density
are contained only in the domain boundary, and hence, no Gibbs
phenomenon arises, if the density decreases sufficiently smooth at the
boundary, e.g., for the $\Gamma=2$ polytropic equation of state. We also
locate the outer boundaries at spatial infinity by employing a radial
coordinate, which is obtained by a transformation $u = 1/r$ in the
outermost domain. With this treatment, the exact boundary conditions at
spatial infinity can be imposed. Because the multidomain method is
employed and field equations are solved for many domains, we split the
field equations into the ``BH part'' and the ``NS part'' like the way
described in Appendix. A of Ref.~\cite{KT_ex}.

Our computational domains centered on the BH are divided into 8 domains
and each of them is covered by $N_r \times N_\theta \times N_\phi = 41
\times 33 \times 32$ collocation points. Similarly, the domains for
the NS are divided into 6 domains and each of them is covered by $N_r
\times N_\theta \times N_\phi = 25 \times 17 \times 16$ collocation
points. Here, $N_r$, $N_\theta$ and $N_\phi$ are the number of
collocation points for radial, polar, and azimuthal directions,
respectively.

\subsection{Iteration procedure} \label{sec:iteration}

Numerical solutions of quasiequilibrium states are obtained by
iteratively solving the basic equations described in
Sec.~\ref{sec:formulation}.  Here, we briefly summarize our procedure of
the iteration. As described in Sec.~\ref{sec:free}, the calculation
should be performed to give correct $M^{\rm NS}_{\rm B}$, $M^{\rm
BH}_{\rm irr}$, and $d$.

First of all, we need to prepare an initial trial solution for the
iterative procedure. For this purpose, we superimpose a Schwarzschild
solution in the isotropic coordinates and a spherical NS. Then, our
iterative procedure is as follows:
\begin{enumerate}
\item Determine the orbital angular velocity $\Omega$ by using
      Eq.~(\ref{eq:omega}).
\item Determine the location of the center of mass (rotational axis) of
      the system.  Because the coordinate separation between two objects
      are initially given, the positions of both objects relative to the
      center of mass are also determined.
\item Solve the equations described in Secs.~\ref{sec:field} and
      \ref{sec:hydro} in each domain.
\item Adjust $M_{\rm P}$ to fix the irreducible mass of the BH to a
      desirable value. After this procedure, determine $M_{\rm \Phi}$ so
      that the condition (\ref{eq:condition_C}) is satisfied.
\item Adjust the maximum enthalpy of the NS, $h_c$, at the center of the
      NS, to fix the baryon rest mass of the NS.
\end{enumerate}
We repeat this procedure until a sufficient convergence is achieved. 
As a measure of the convergence, we monitor 
the relative difference between the enthalpy field of
successive steps. Typically, we stop the iteration when the 
following condition is satisfied:
\beqn
\sum_{i,j,k} |1-h^n_{i,j,k}/h^{n-1}_{i,j,k}| \leq 10^{-6},
\eeqn
where $n$ denotes the iteration step, and $(i, j, k)$ the 
collocation points of $(r, \theta, \varphi)$. 

\section{Numerical results} \label{sec:results}

\begin{figure*}[t]
\includegraphics[width=80mm,clip]{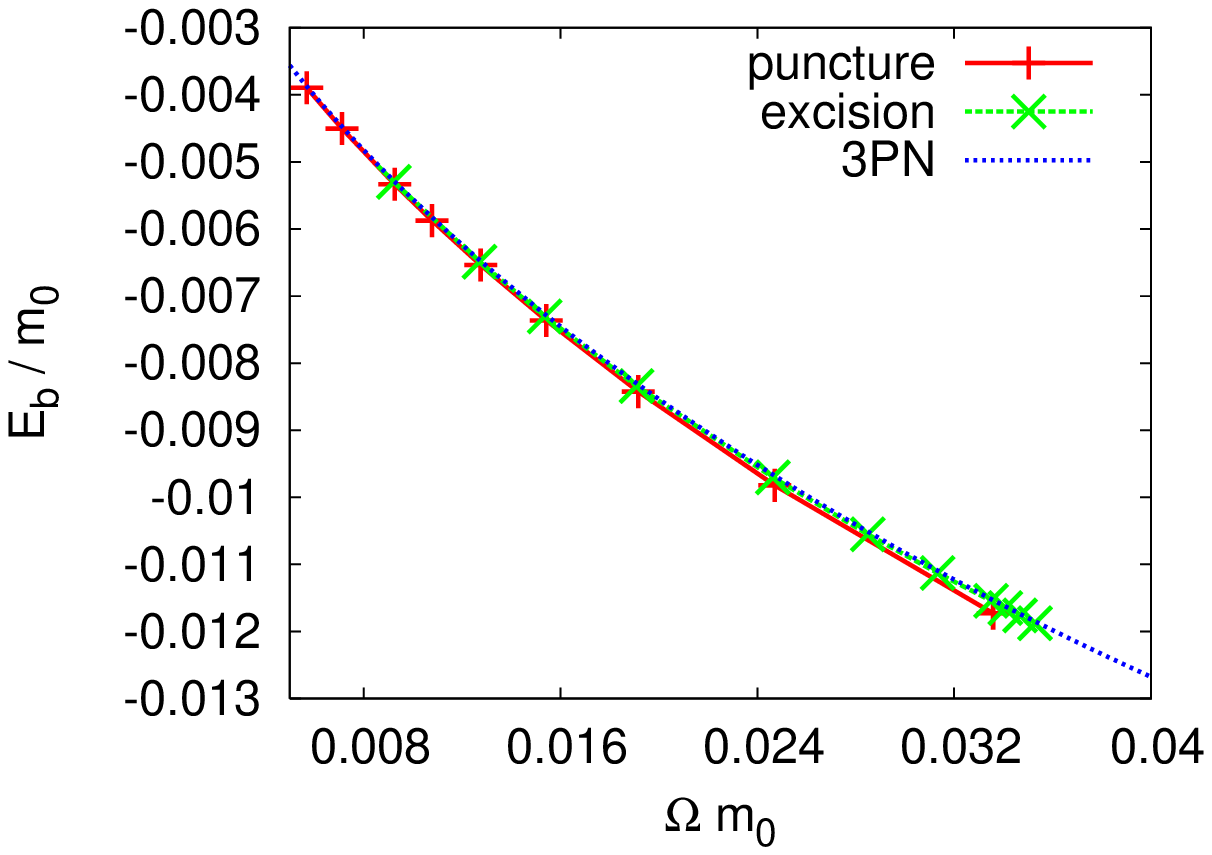}
\includegraphics[width=80mm,clip]{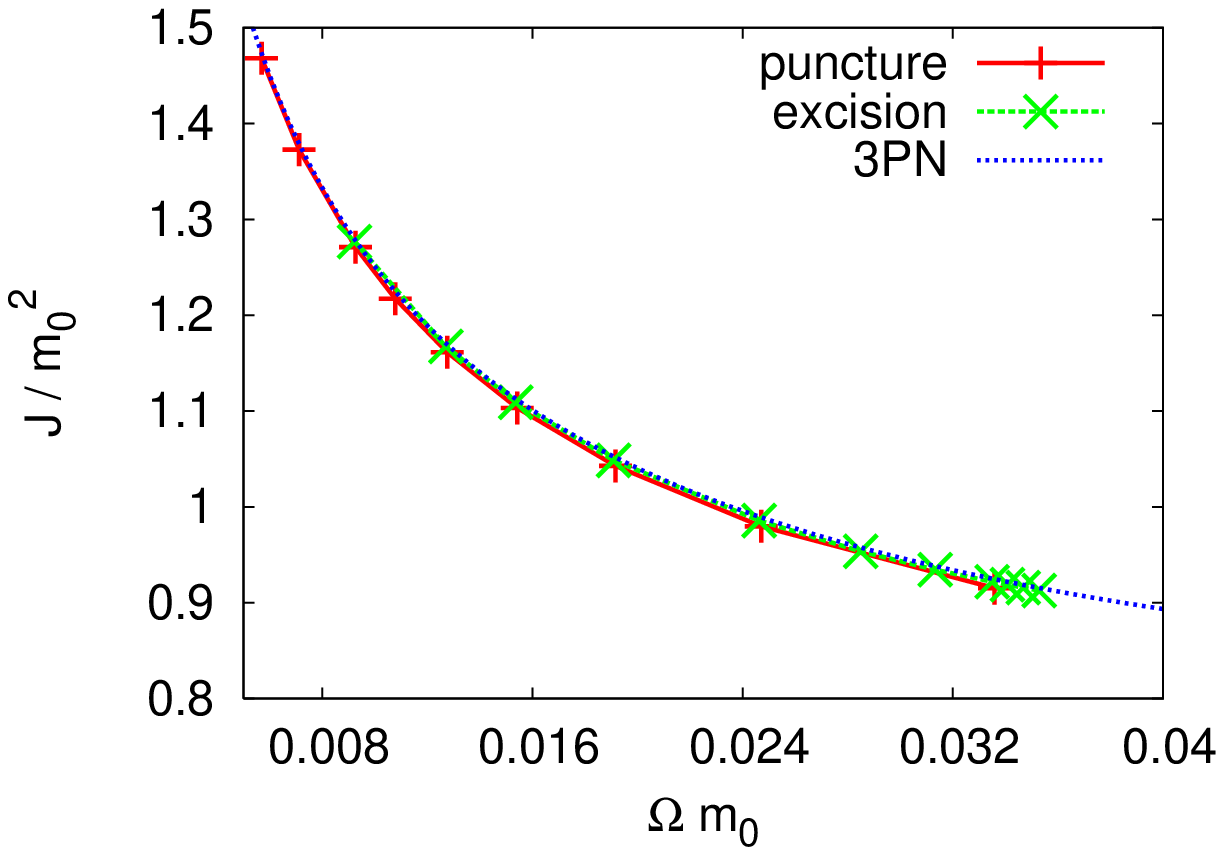}\\
\includegraphics[width=80mm,clip]{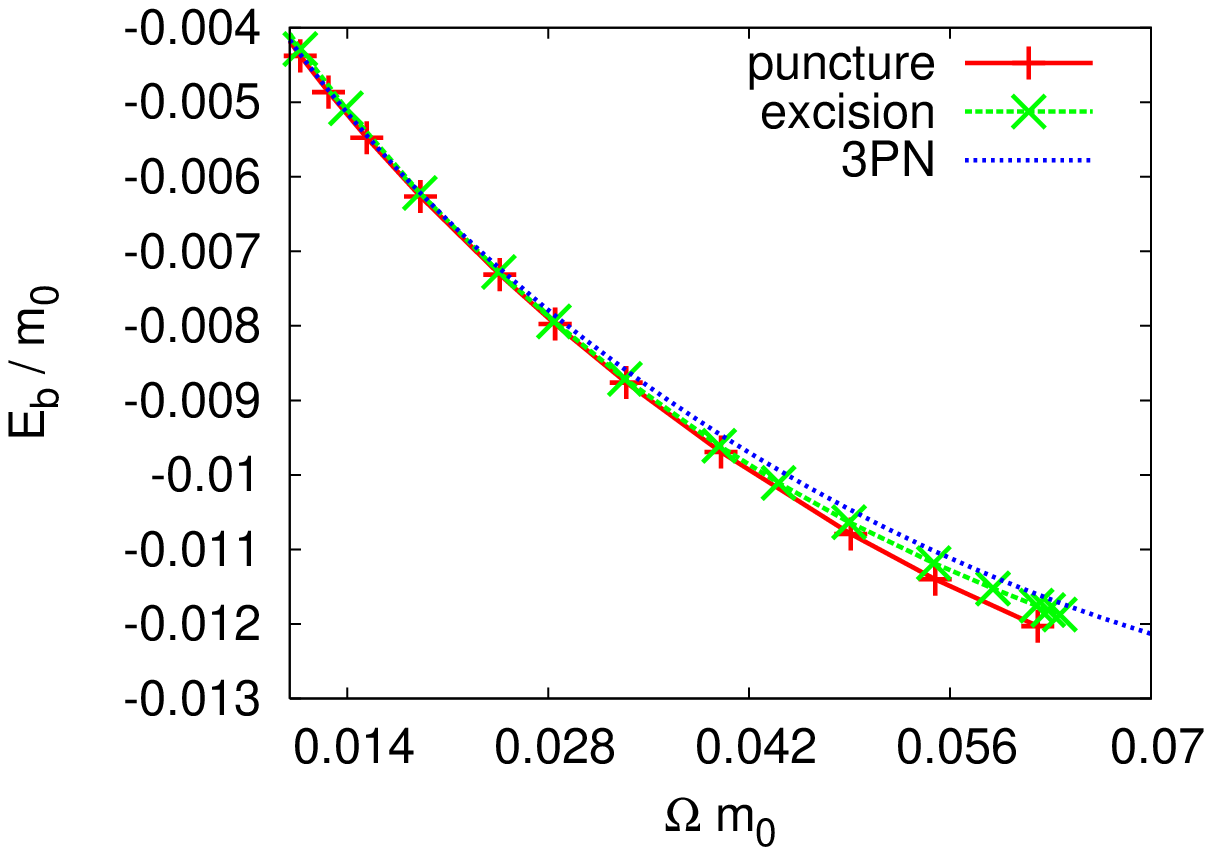}
\includegraphics[width=80mm,clip]{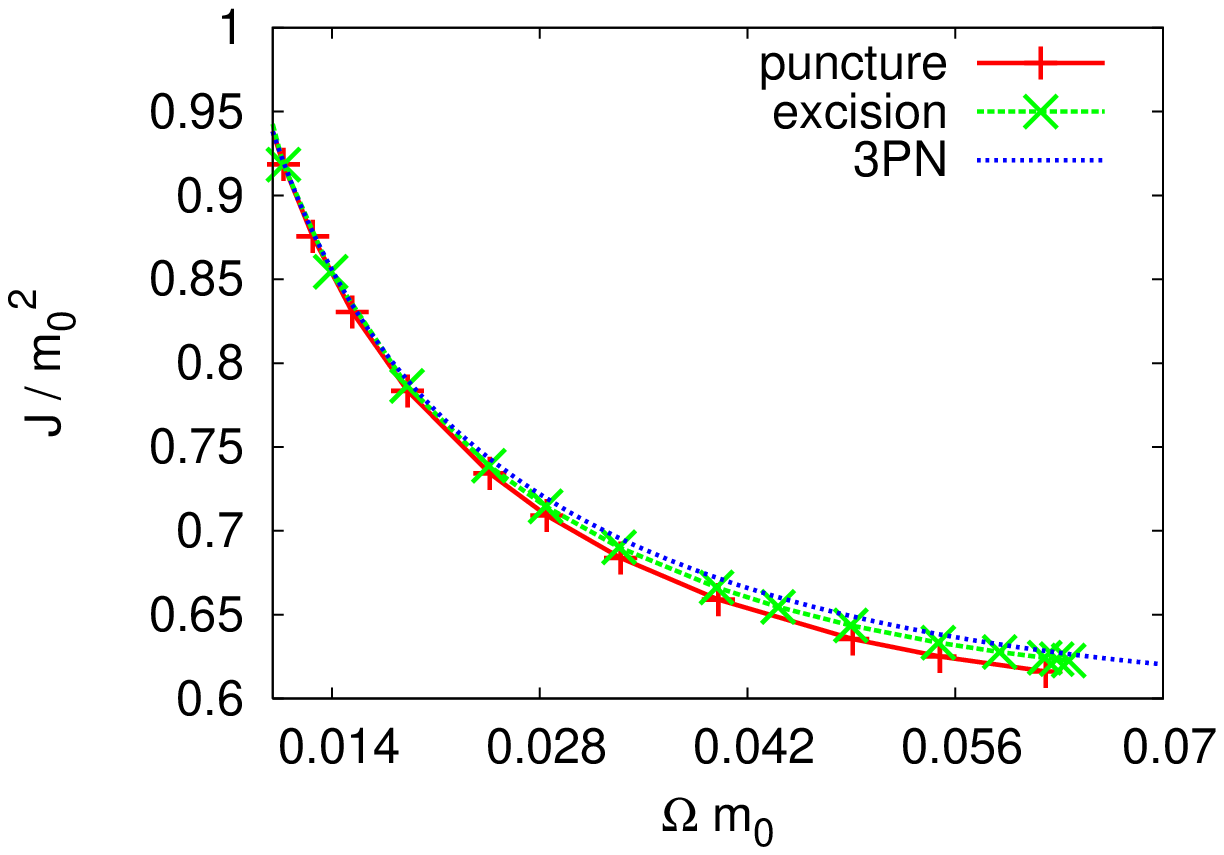}\\
\includegraphics[width=80mm,clip]{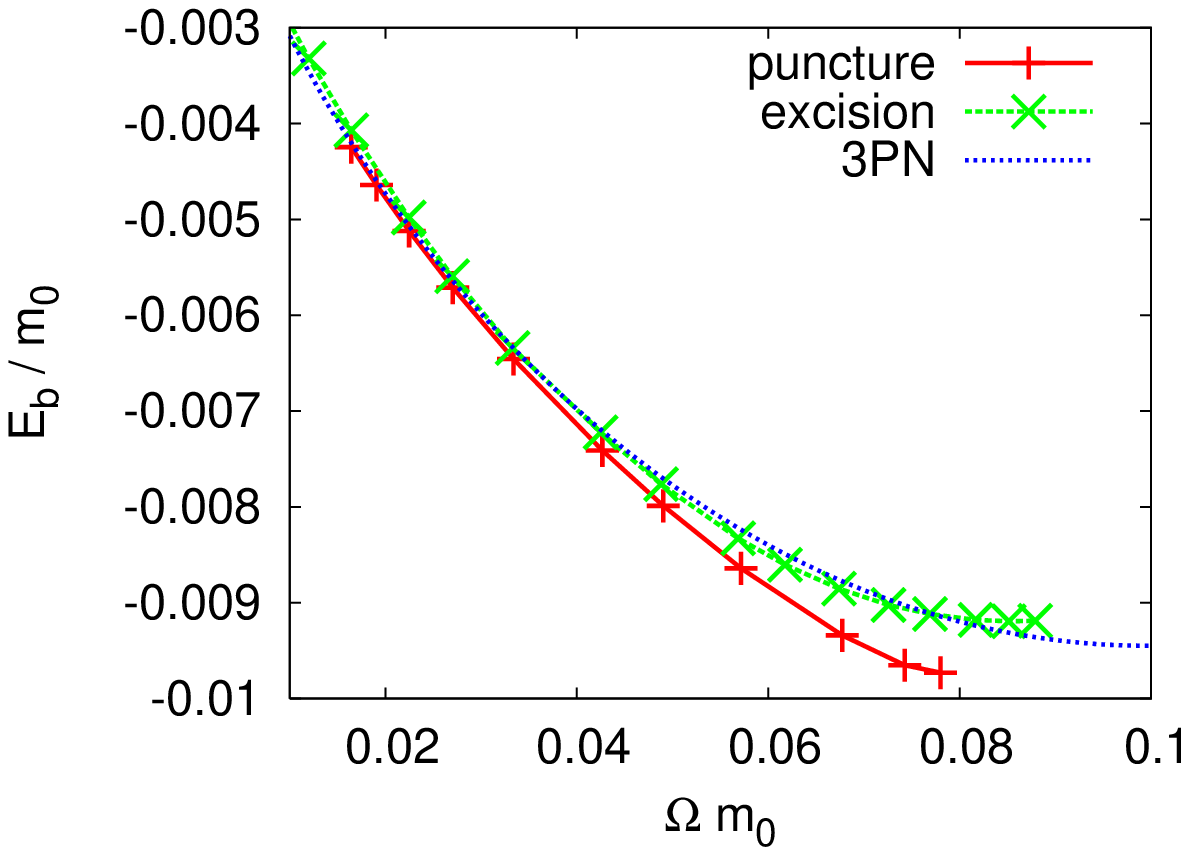}
\includegraphics[width=80mm,clip]{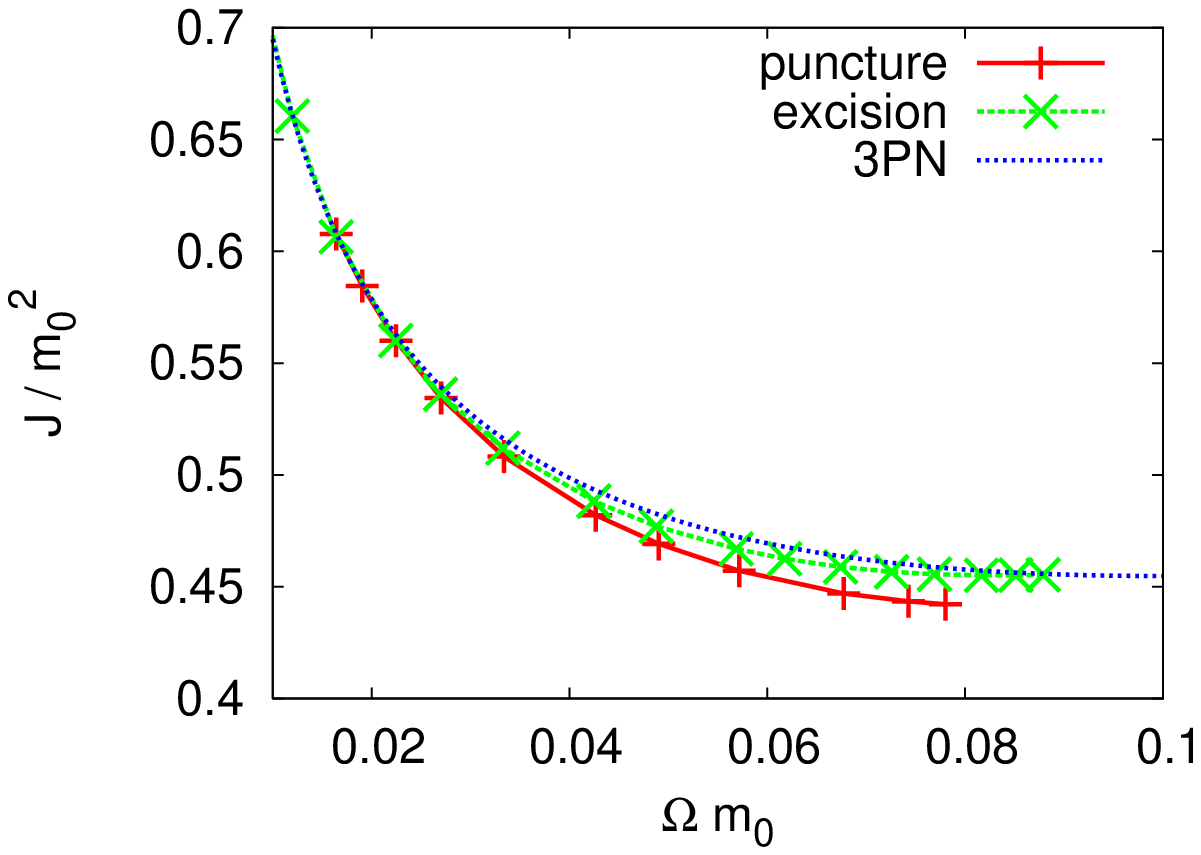}
\caption{Left panels: Binding energy $E_b/m_0$ as a function of
$\Omega m_0$ for ${\bar M}^{\rm NS}_{\rm B}=0.15$ and $Q=1$ (upper
panel), 3 (middle panel), and 5 (lower panel). The solid (red) and
dashed (green) curves show the results obtained in the moving-puncture
method with the $\beta^{\varphi}$ condition and in the excision method
\cite{KT_ex2}, respectively. The dotted (blue) curve denotes the
result in the 3PN approximation \cite{3PN}.  Right panels: The same as
 the left panels but for the total angular momentum $J/m^2_0$ as a
 function of $\Omega m_0$.}
\label{fig:energy1R1}
\end{figure*}

\begin{figure*}[t]
\includegraphics[width=80mm,clip]{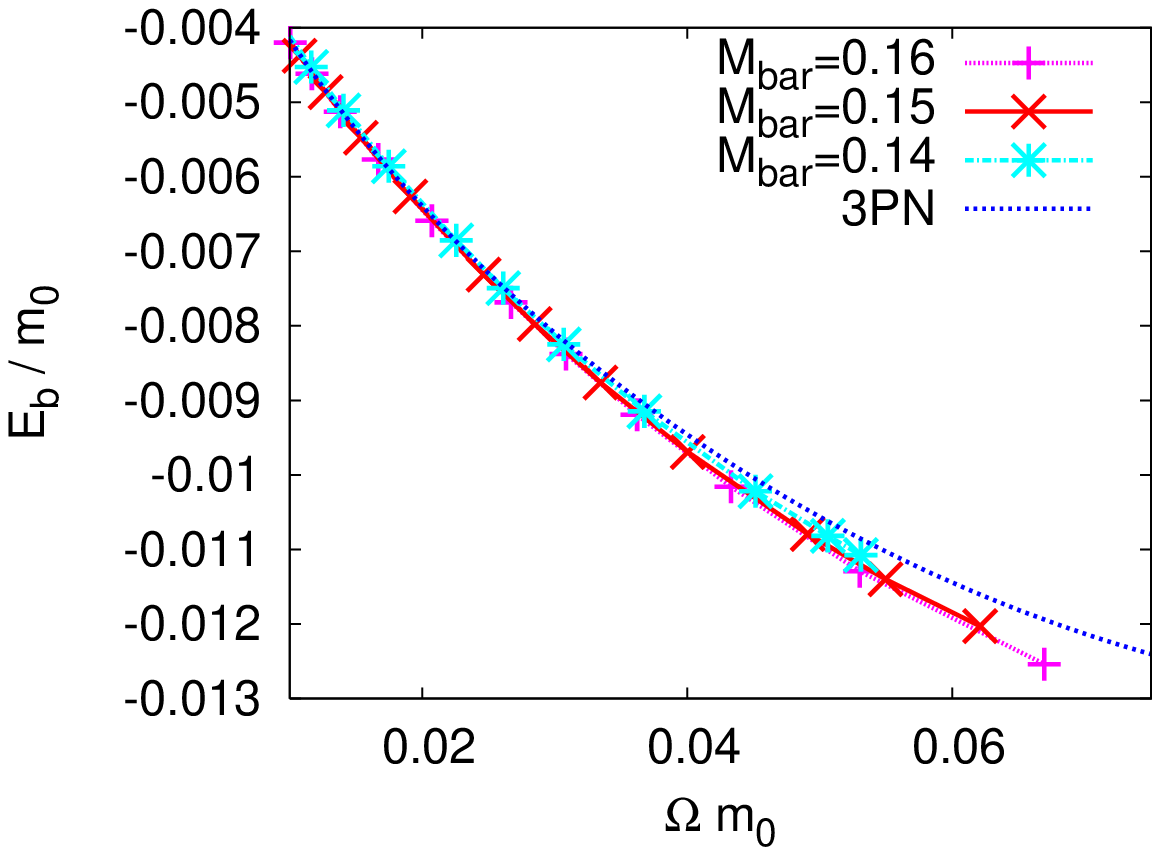}
\includegraphics[width=80mm,clip]{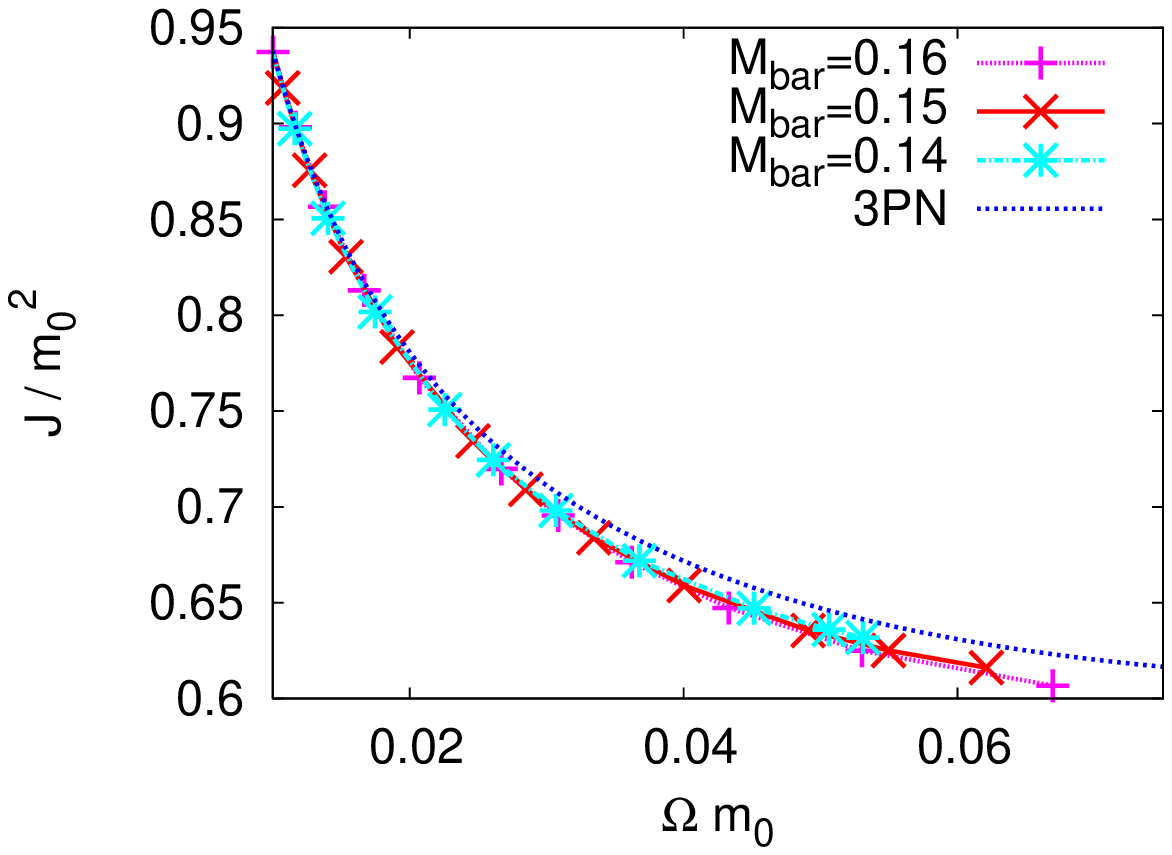}
\caption{The same as Fig.~\ref{fig:energy1R1} but for ${\bar M}^{\rm
 NS}_{\rm B}=0.14$, 0.15, and 0.16, and for $Q=3$ in the moving-puncture
 method.}
\label{fig:ener3R1comp}
\end{figure*}

Throughout this paper, we characterize a quasiequilibrium sequence by
two parameters; the baryon rest mass of the NS, $\bar{M}^{\rm NS}_{{\rm
B}}$, and the ratio of the irreducible mass of the BH to the ADM mass of
the NS in isolation, $Q \equiv M^{\rm BH}_{\rm irr}/M^{\rm NS}_{{\rm
ADM},0}$. We focus on the sequences of ${\bar M}^{\rm NS}_{\rm B}=0.14$,
0.15, and 0.16 following Ref.~\cite{KT_ex2}.  For these cases, the
compactness of the NSs is ${\cal C}=0.1321$, 0.1452, and 0.1600,
respectively, implying that we choose moderately large values for the
compactness. Note that the maximum value of $\bar{M}^{\rm NS}_{{\rm B}}$
is about 0.18 for the $\Gamma=2$ polytropic equation of state. At that
value of $\bar{M}^{\rm NS}_{{\rm B}}$, the compactness is about ${\cal
C}=0.21$. Here, the compactness is defined by
\begin{equation}
{\cal C} \equiv \frac{M^{\rm NS}_{{\rm ADM},0}}{R_0}, 
\end{equation}
where $R_0$ is the circumferential radius of the NS in isolation. For
the mass ratio, we choose $1 \le Q \le 5$. 


\subsection{Binding energy and total angular momentum in 
the $\beta^{\varphi}$ condition} \label{sec:energy}

Figure~\ref{fig:energy1R1} plots the binding energy and total angular
momentum as functions of $\Omega m_0$ for $Q=1$, 3, 5, and ${\bar
M}^{\rm NS}_{\rm B}=0.15$ in the $\beta^\varphi$ condition.  For
comparison, the results in the excision approach \cite{KT_ex2} and in
the 3PN approximation are also plotted. 

For large orbital separations (small values of $\Omega m_0 \alt 0.02$),
the results obtained in the moving-puncture method (with the
$\beta^{\varphi}$ condition) agree well with those derived by the 3PN
approximation and in the excision approach \cite{KT_ex2} irrespective of
the mass ratio. For smaller separations, however, the degree of
agreement among three results depends on the mass ratio. For $Q=1$,
three results agree within $\sim 1\%$ error. By contrast, for $Q=3$ and
5, a significant deviation of order $\sim 10\%$ arises among three
results for $\Omega m_0\agt 0.03$. In particular, for $Q=5$, the results
in the moving-puncture approach (with the $\beta^{\varphi}$ condition)
disagree significantly with other two results (see lower panels of
Fig.~\ref{fig:energy1R1}). Because the tidal effect does not play an
important role and the orbital velocity is at most $\sim 0.1c$ for a
fairly distant orbit of $\Omega m_0 \sim 0.03$, the numerical results
should agree with the results in the 3PN approximation, at least
approximately. This implies that the quasiequilibrium state obtained in
the moving-puncture approach (with the $\beta^\varphi$ condition) is not
in a quasicircular orbit for the close orbit of $\Omega m_0 \gtrsim
0.03$ and for $Q \ge 3$; it would be an eccentric orbit. Also,
quasiequilibrium states in this moving-puncture approach appear to be
inferior to those in the excision approach in that they show systematic
deviations from other two results.

One possible reason for this deviation may stem from the condition for
determining the center of mass of the system. This point is explored in
detail in Sec.~\ref{sec:CM}. Another possible reason is that the BH
might have nonzero spin in the present approach. Indeed, in the excision
framework, it has been found that the ``leading-order approximation''
leads to a slightly spinning BH \cite{cookpheiffer,caudill}. To obtain a
strictly nonspinning BH, a computation has to be performed with an
improved method for determining the spin angular velocity of the
BH. Motivated by this fact, we measured the spin of the BH using a
method proposed by Cook and Whiting \cite{CW2007}. However, we find that
the BH has negligible spin of order $S/M^2_{\rm irr} \lesssim 10^{-5}$,
and therefore, the deviation between the results in the moving-puncture
method (with the $\beta^{\varphi}$ condition) and others is not caused
by the spin of the BH.

Numerical results of binding energy and total angular momentum in the
moving-puncture method as a function of $\Omega m_0$ for different
compactness of the NS are plotted in Fig.~\ref{fig:ener3R1comp}. This
shows that the feature of the results described above holds irrespective
of the compactness of the NS.  For smaller values of the compactness
(i.e., for smaller values of $\bar M^{\rm NS}_{\rm B}$), the binding
energy and the angular momentum are slightly larger for a given value of
$\Omega m_0\agt 0.04$. This is due to the fact that for less compact
NSs, the tidal-deformation effect on the NS plays an important role in
increasing these quantities in close orbits \cite{lrs1993}. We note
that, in the moving-puncture approach, it is possible to obtain the
sequences of quasiequilibria for the NSs with the compactness ${\cal C}
\lesssim 0.18$ for the $\Gamma = 2$ polytropic equation of
state. Meanwhile, our computations do not show adequate convergence for
the binaries containing more compact NSs, because such NSs are close to
the most compact ones allowed by the given equation of state, ${\cal C}
\sim 0.21$ in this case, and are difficult to compute accurately. For
the $\Gamma > 2$ equation of state for which the maximum compactness is
larger than 0.21, we are able to obtain NSs of compactness $\sim 0.2$.

Before closing this subsection, we point out the following issue found
from Fig.~\ref{fig:energy1R1}: The angular momentum for a given value
of $\Omega m_0$ in the numerical results is always smaller than that in
the 3PN approximation for $\Omega m_0\agt 0.03$.  This holds for the
results not only in the moving-puncture approach (with the
$\beta^\varphi$ condition) but also in the excision approach. For a
fairly distant orbit of $\Omega m_0 \sim 0.03$, the 3PN approximation
should be an excellent approximation of general relativity, and hence,
provides a highly accurate result.  This implies that angular momentum
in the numerical results is smaller than that for the real quasicircular
state. As shown in Sec.~\ref{sec:simulation}, this is indeed the
case. Figure \ref{fig:energy1R1} indicates that for numerical
simulation, these quasiequilibria may not be good initial conditions,
and an improved quasiequilibrium would be favorable as the initial
condition of numerical simulation.

\subsection{Effect of the center of mass} \label{sec:CM}

\begin{figure*}[t]
\includegraphics[width=80mm,clip]{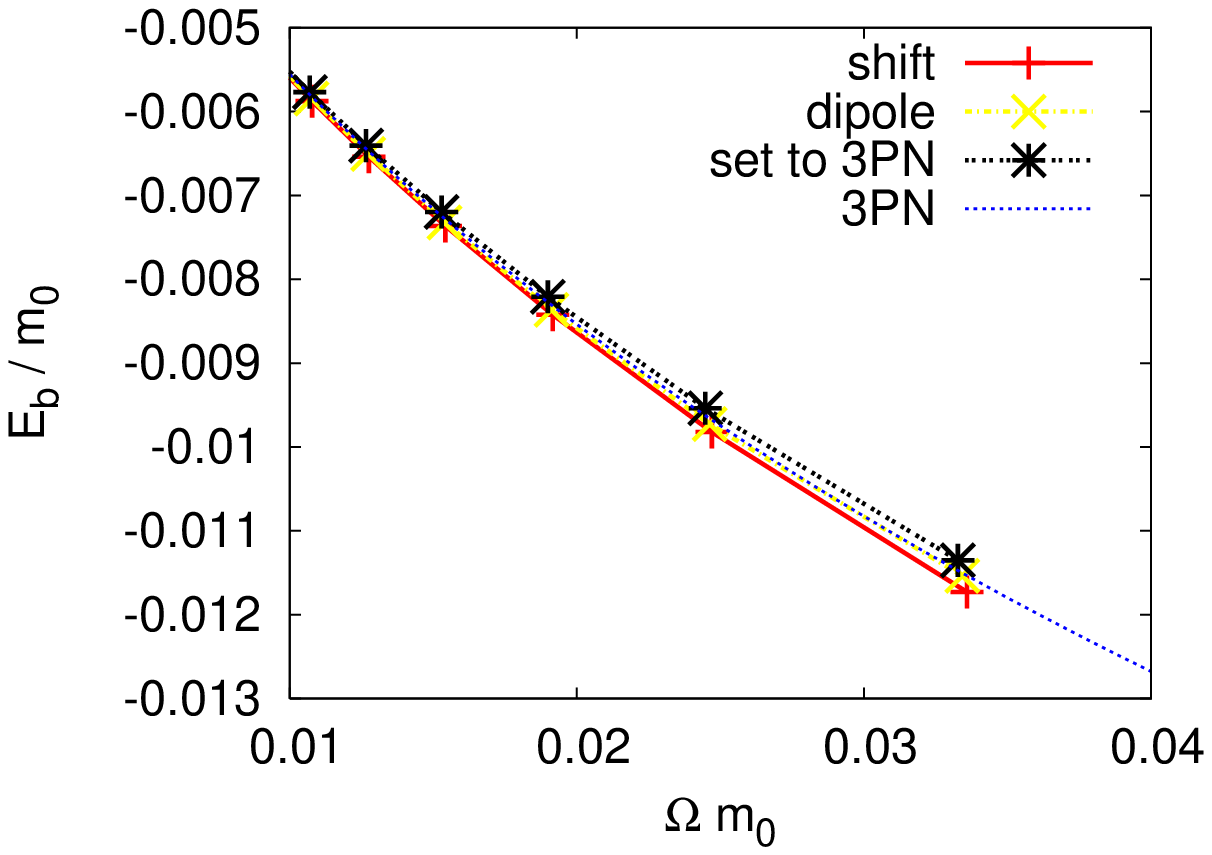}
\includegraphics[width=80mm,clip]{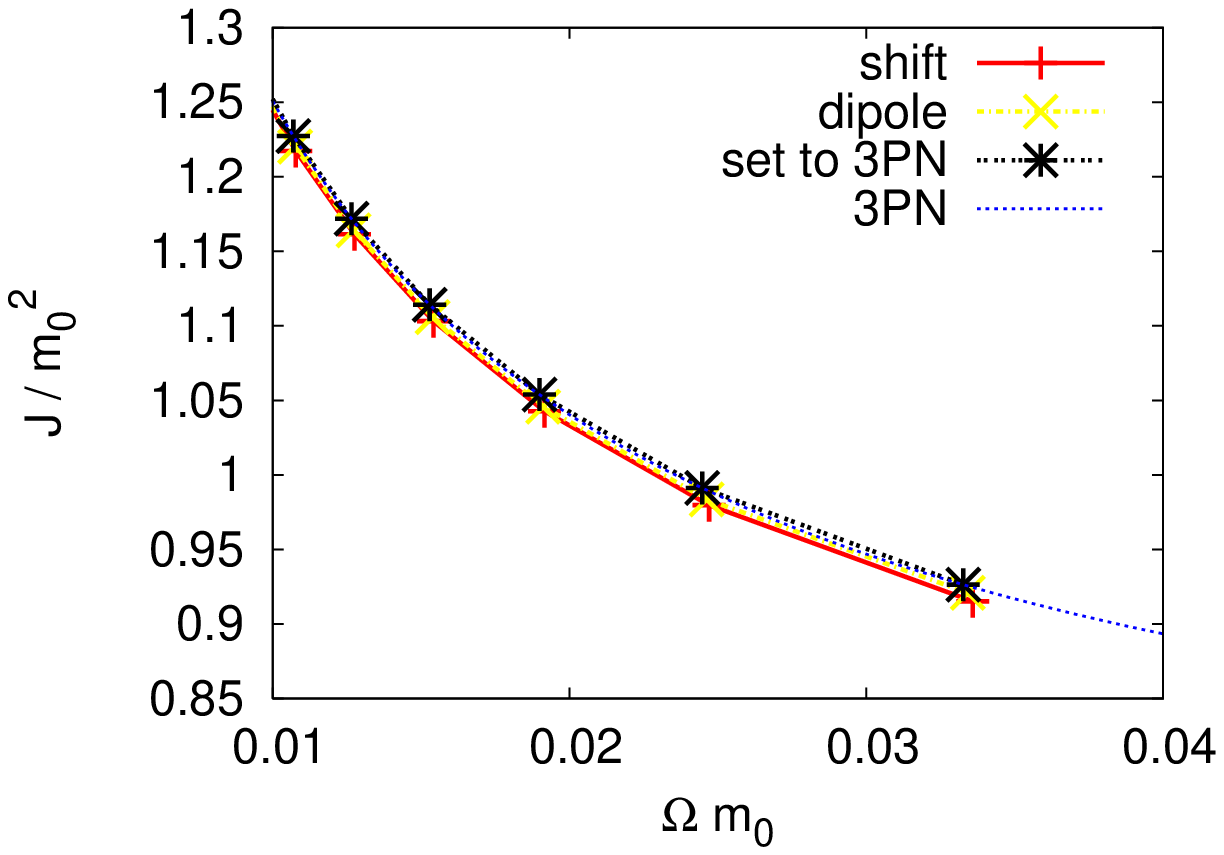} \\
\includegraphics[width=80mm,clip]{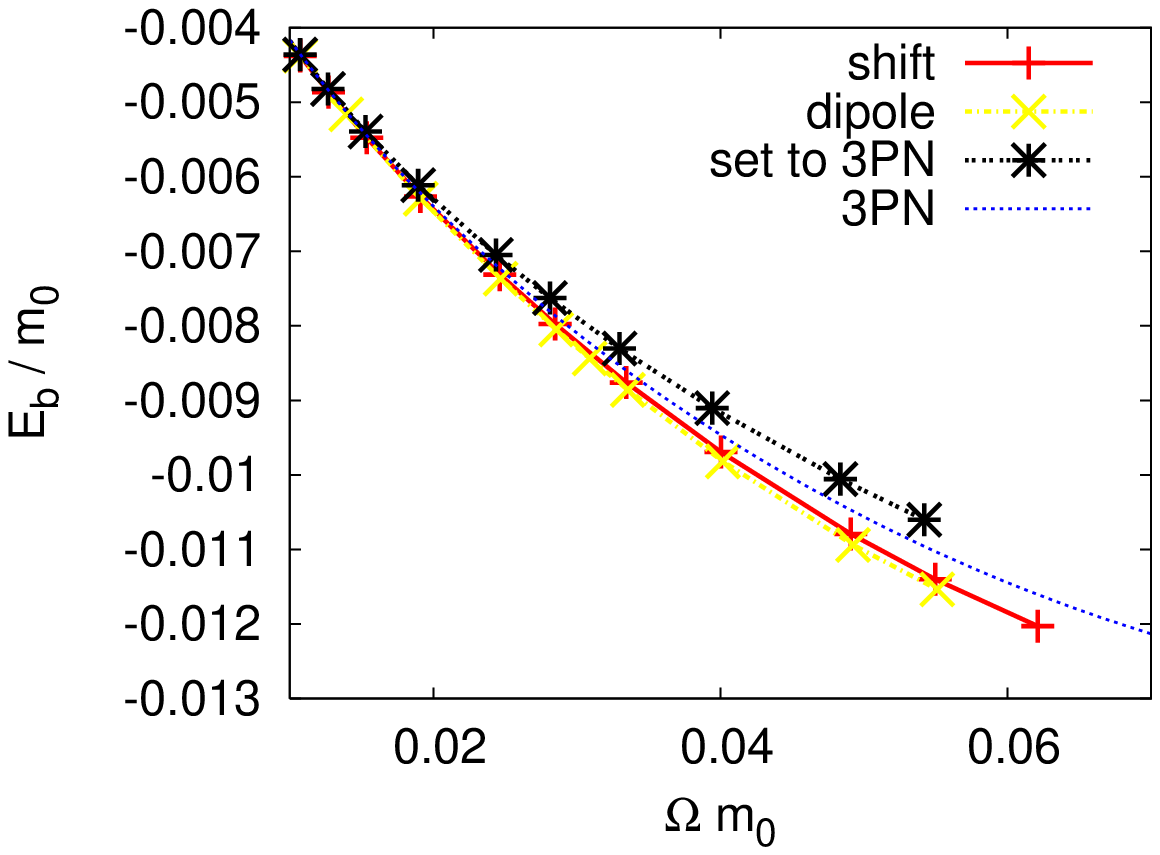}
\includegraphics[width=80mm,clip]{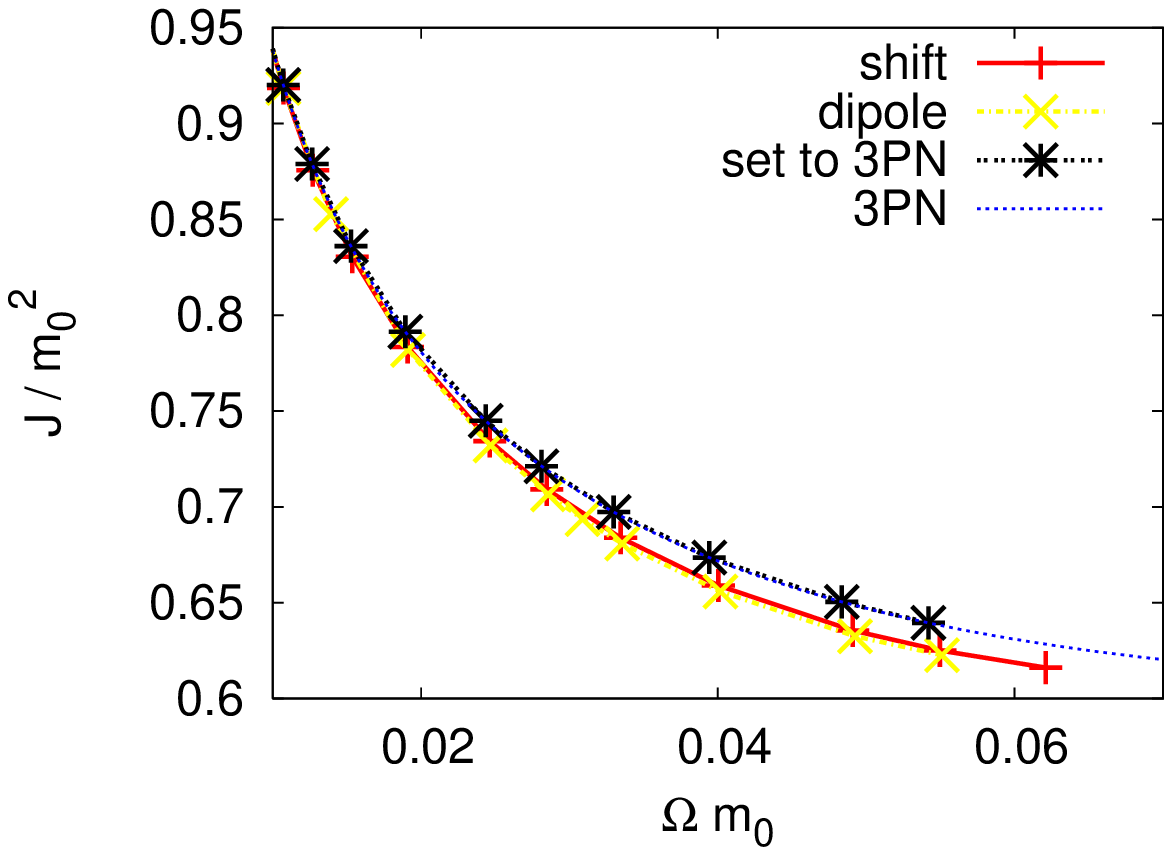} \\
\includegraphics[width=80mm,clip]{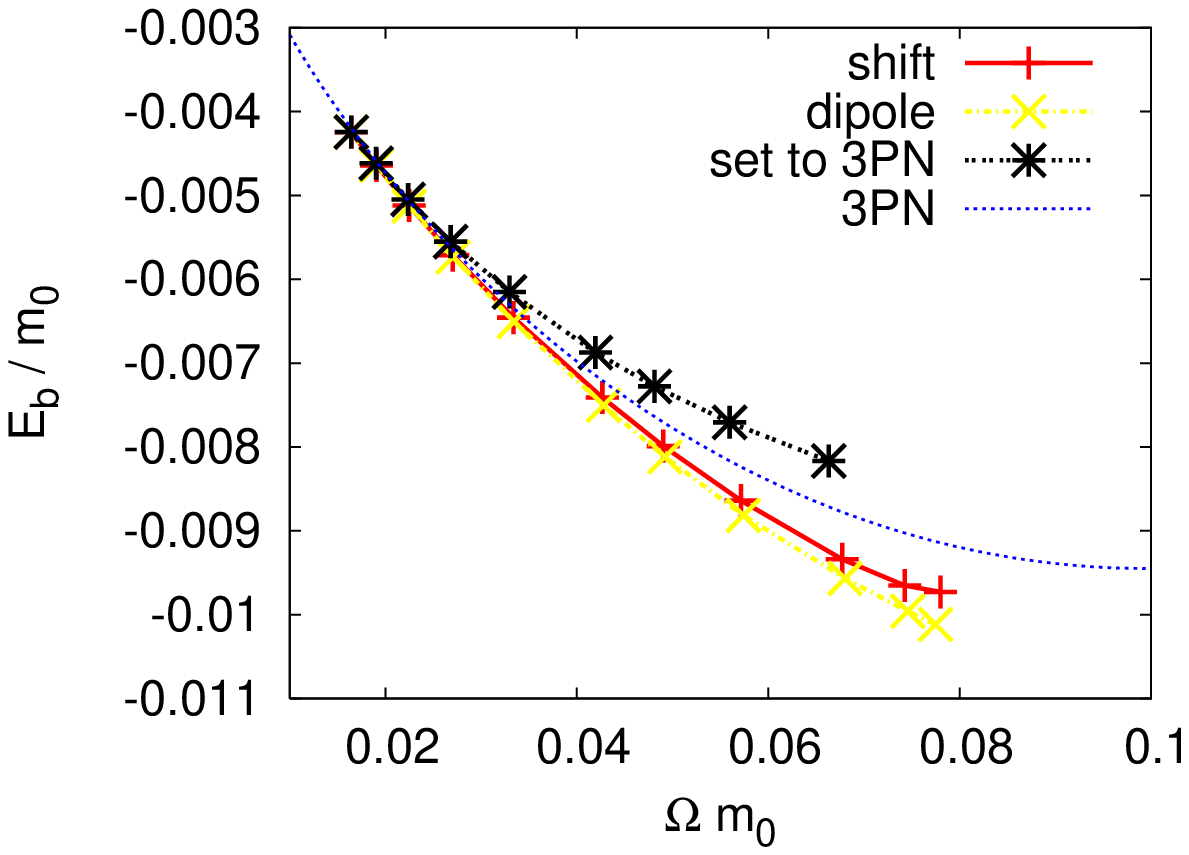}
\includegraphics[width=80mm,clip]{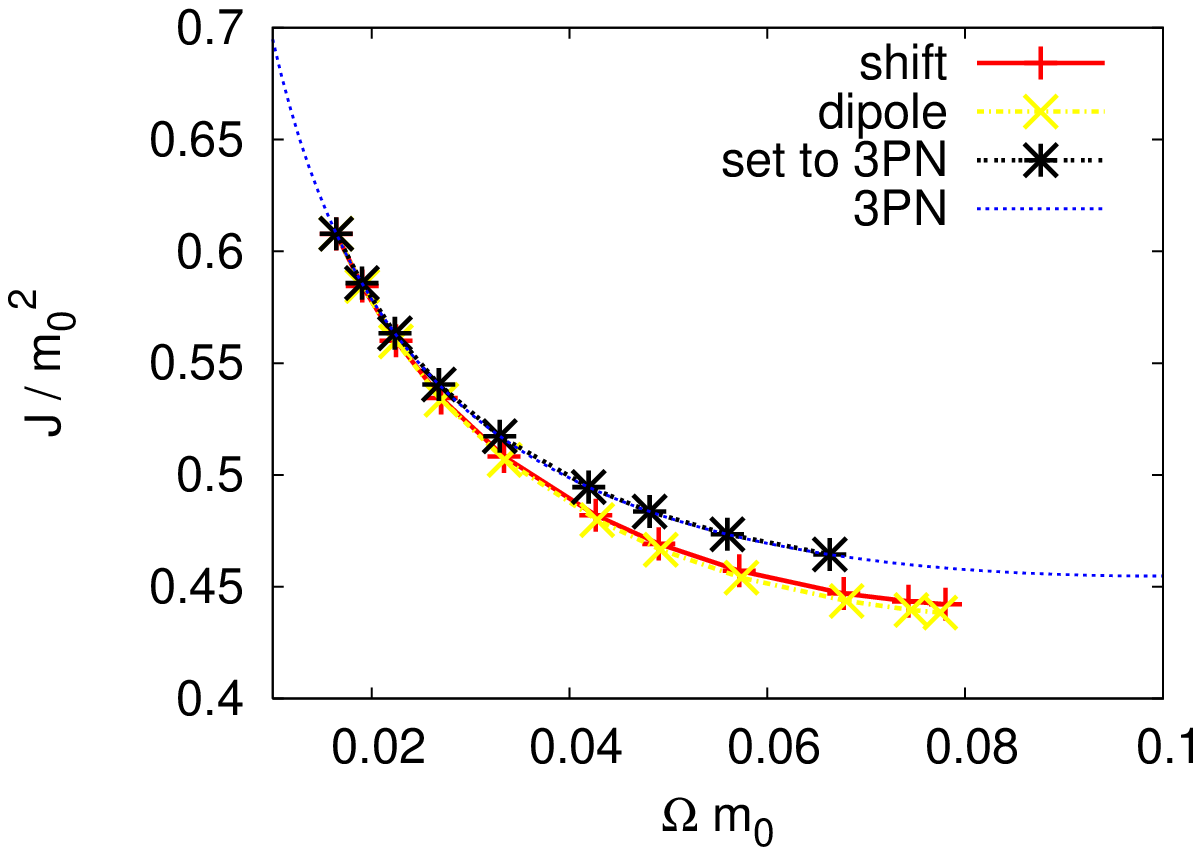}
\caption{Left panels: Binding energy $E/m_0$ as a function of $\Omega
m_0$ for ${\bar M}^{\rm NS}_{\rm B}=0.15$ and $Q=1$, 3, and 5 in the
moving-puncture approach with three different conditions for
determining the center of mass of system. The dotted curve denotes the
result in the 3PN approximation.  Right panels: The same as the left
 panels but for the total angular momentum $J/m^2_0$ as a function of
 $\Omega m_0$.}
\label{fig:rotener1R1}
\end{figure*}

As mentioned in Sec.~\ref{sec:free}, we have no definitive guidance for
determining the position of the center of mass in the moving-puncture
framework. In the excision framework, the position is automatically
determined so that the ADM linear momenta should vanish. By contrast, in
the moving-puncture framework, this condition was already used to
determine another free parameter, $P_i^{\rm BH}$.

As described in Sec.~\ref{sec:free}, we have at least three methods for
determining the center of mass, and numerical results depend strongly on
them. In this subsection, we compare these numerical results. 

In Fig.~\ref{fig:rotener1R1}, we show sequences of $\bar{M}^{\rm
NS}_{\rm B}=0.15$ for $Q=1$, 3, and 5 obtained by three different
methods for the center of mass. ``shift,'' ``dipole,'' and ``set to
3PN'' denote the results derived in the $\beta^{\varphi}$, dipole, and
3PN-J conditions, respectively.

The numerical results in the $\beta^\varphi$ and dipole conditions show
a similar behavior. For $Q=1$, these results agree approximately with
those in the 3PN approximation, and for $Q=3$ and 5, the deviation from
the 3PN results becomes significant as pointed out in
Sec.~\ref{sec:energy}. The deviation from the 3PN results is slightly
smaller for the results in the $\beta^\varphi$ condition for $Q=3$ and 5
than those in the dipole condition. This indicates that the
$\beta^\varphi$ condition is slightly better for computing unequal-mass
BH-NS binaries in quasiequilibrium, although the deviation from the 3PN
results is larger than 1\% for $\Omega m_0\agt 0.03$.

The sequence obtained in the 3PN-J condition shows rather different
behavior. By definition, the angular momentum agrees with the results
obtained in the 3PN approximation. As a result, however, the binding
energy becomes larger than that in the 3PN approximation, in contrast to
the results in other two methods. A possible interpretation for the
excess of the binding energy is that junk gravitational radiation or
nonstationary kinetic energy (e.g., oscillation of a nonstationary BH)
are included in the data. However, if these nonstationary components are
radiated away during numerical evolution, this initial data could
provide an approximate quasicircular state. 

\subsection{Assessing quality of quasiequilibrium by numerical simulation}
\label{sec:simulation}

The results presented in the previous subsection show that the
quasiequilibrium states computed in three conditions of the
moving-puncture approach are not in quasicircular orbits for $Q \not=1$, 
but rather are likely to be in eccentric orbits. However, if the
eccentricity is small enough, it quickly approaches zero during
evolution, resulting in that a realistic binary, i.e., a (n
approximately) quasicircular state, is provided. To assess circularity
of quasiequilibrium states as the initial condition of numerical
relativity, we performed a numerical simulation, choosing a binary of
$\bar M_{\rm B}^{\rm NS}=0.15$ and $Q=3$ and 5. In this subsection, we
present the results of the numerical simulation. In this experiment, we
adopt the initial data obtained in the $\beta^\varphi$ and 3PN-J
conditions of the moving-puncture approach. For all the data, the
angular velocity is chosen to be $\Omega m_0\approx 0.033$.  The
numerical simulation was performed using our code {\sc sacra}, in which
an adaptive mesh refinement algorithm is implemented. The formulation,
gauge condition, and numerical scheme adopted in {\sc sacra} are the
same as described in Ref.~\cite{SACRA}.

\begin{figure*}[th]
\includegraphics[width=85mm,clip]{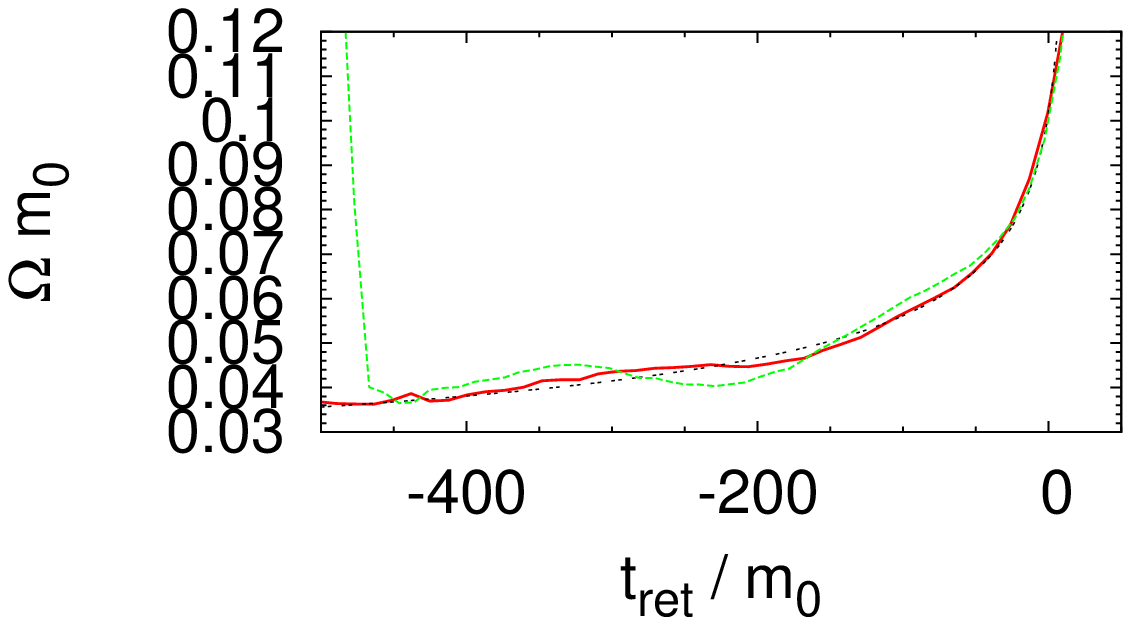}
\includegraphics[width=85mm,clip]{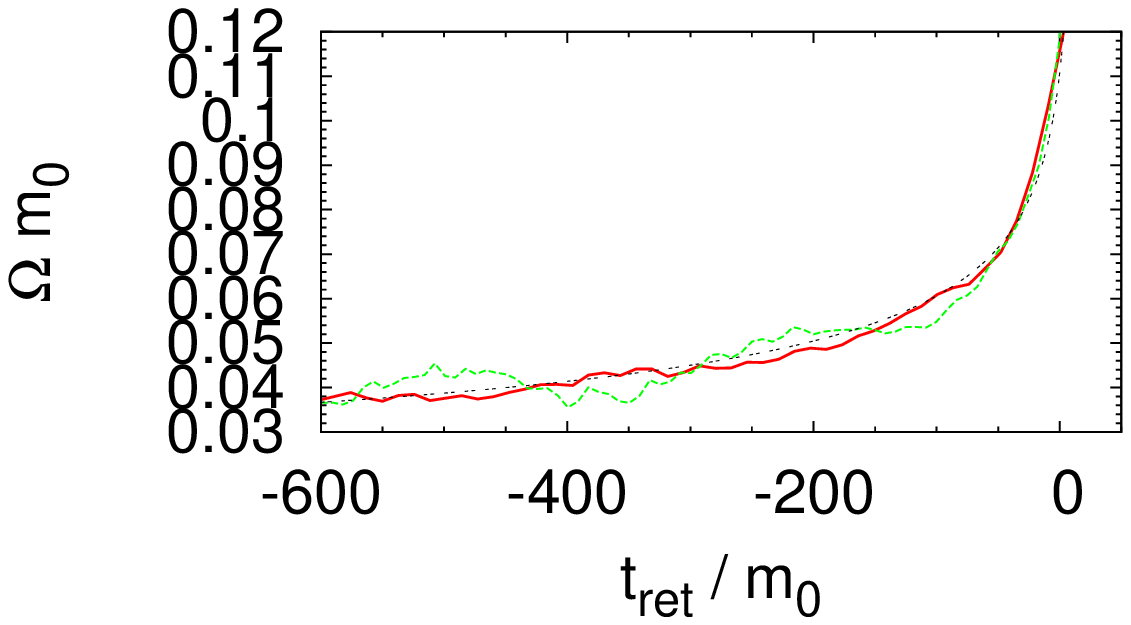}
\vspace{-5mm}
 \caption{Angular velocity of gravitational waveforms as a function of
 retarded time, $t_{\rm ret}$, for $Q=3$ (left) and $Q=5$ (right) 
and for ${\cal C}=0.15$. $t_{\rm ret}=0$ approximately denotes
 the merger time. The solid and dashed curves denote the results in the
 3PN-J condition and  the $\beta^\varphi$ condition, respectively. The
 dot-dotted curve is the result in the 3PN approximation (Taylor T4
 formula; e.g., Ref.~\cite{BHBH28}).}
 \label{fig:w3}
\end{figure*} 

Figure~\ref{fig:w3} plots evolution of the orbital angular velocity as a
function of retarded time. Here, the angular velocity is calculated by
the quadrupole mode of gravitational waveforms by \cite{BHBH23}
\beqn
\Omega(t)
={1 \over 2} {|\Psi_4(l=m=2)| \over \displaystyle \Big|\int dt
\Psi_4(l=m=2)\Big|},
\label{gwangv}
\eeqn 
where $\Psi_4$ denotes the outgoing part of the Newman-Penrose
quantity. This figure illustrates that the orbit of the binaries is
eccentric for all the cases, reflecting the fact that the circular orbit
is not provided initially. For the initial data obtained in the
$\beta^\varphi$ condition, the eccentricity
appears to be of order 0.1. Moreover, the eccentricity does not reduce
to zero even at the onset of merger.  We note that for these cases, the
binary spends in the inspiral phase for 4--5 orbits before the onset of
the merger. Nevertheless, the eccentricity is not sufficiently reduced
by the gravitation radiation reaction and a quasicircular orbit is not
achieved before the merger. The reason for this is that the initial
eccentricity is too large.

By contrast, for the initial data obtained in the 3PN-J condition, the
eccentricity appears to be much smaller than those for other two
cases. This suggests that the initial eccentricity is smaller. 

By comparing the numerical results with that obtained in the 3PN
approximation (dot-dotted curve), we find that the modulation
amplitude of the angular velocity is $\Delta (\Omega m_0) \approx
0.004$--0.006 even just before the merger (at $\Omega m_0\sim 0.05$)
in the $\beta^\varphi$ condition. Eccentricity of the orbit is
approximately estimated to be $2\Delta \Omega/3\Omega$ for a slightly
eccentric orbit. Thus, the eccentricity for these cases is $\sim
0.05$--0.08.

By contrast, the modulation amplitude is at most $\Delta(\Omega m_0)
\approx 0.003$ for the initial data given in the 3PN-J
condition. Furthermore, the modulation is suppressed to be $\alt 0.001$
just before the onset of merger ($\Omega m_0\agt 0.05$) in this case,
indicating that the eccentricity is $\alt 0.01$.  This shows that the
moving-puncture approach with the 3PN-J condition is superior for
providing the initial data for the numerical-relativity simulation.

\subsection{Mass-shedding limit} \label{sec:tidal}

As analyzed in detail in Refs.~\cite{KT_ex,KT_ex2}, the NS is strongly
subject to tidal deformation by a companion BH outside their innermost
stable circular orbit (ISCO), if the mass ratio of the system is small
enough or the radius of the NS is large enough. In the case that the
tidal field of the BH is strong enough, the Lagrangian point enters
inside the surface of the NS. Then the NS cannot be in equilibrium any
longer because mass shedding occurs from the inner edge of the NS's
surface. We here determine the condition for the onset of mass shedding
in the moving-puncture approach with the $\beta^{\varphi}$ condition. As
shown in the previous two subsections, the quasiequilibrium states
obtained in this approach are not quasicircular states but slightly
eccentric ones. This seems to be also the case for the quasiequilibrium
states obtained in the excision approach employed in
Refs.~\cite{KT_ex,KT_ex2}. Even though there exists nonzero eccentricity
in the data, we think that it still deserves to reinvestigate the
mass-shedding limit in the moving-puncture approach and to compare the
results with those obtained in the excision approach.

In the spectral methods one cannot determine any equilibrium
configuration for a star with irregular surface shape (i.e., with a
cusp). At the onset of mass shedding, the inner edge of the NS has a
cusp, and hence, it is not possible to accurately compute
quasiequilibrium states of the NS in close orbits with a BH. Thus, we
infer the orbital separation (or angular velocity) at the onset of mass
shedding from quasiequilibrium states of slightly more distant orbits
than the mass-shedding configuration has.  For this procedure, we define
a mass-shedding indicator $\chi$ \cite{irrot_hydro} as the fraction of
the radial derivative of the enthalpy at the NS surface,
\begin{equation}
 \chi \equiv \left(\frac{(\partial(\ln h)/\partial
  r)_{\rm eq}}{(\partial(\ln h)/\partial r)_{\rm pole}}\right),
\end{equation}
where subscripts ``eq'' and ``pole'' imply that the partial derivative
is taken with respect to the radial direction connecting the centers of
the NS and BH in the equatorial plane and along the polar direction,
respectively.  For the infinite separation, the NS becomes spherical and
$\chi=1$, whereas $\chi$ decreases with increasing the degree of tidal
deformation and eventually becomes zero at the onset of mass
shedding. Because our code does not converge to give quasiequilibrium
for $\chi \rightarrow 0$, we first compute a sequence of $\chi$ for
close orbits, and then, derive a fitting formula for the curve of $\chi$
as a function of orbital separation. By using this fitting formula, we
determine the orbital separation for $\chi=0$.

\begin{figure}[th]
 \includegraphics[width=80mm,clip]{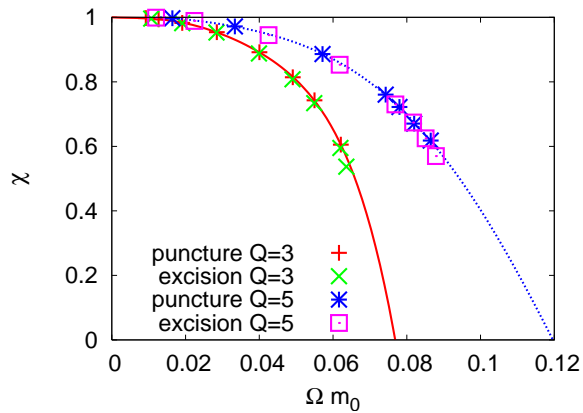}
\caption{The mass-shedding parameter $\chi$ versus $\Omega m_0$ for
 the sequences of ${\bar M}^{\rm NS}_{\rm B}=0.15$, and $Q=3$ and 5. We
 also show the results obtained in the excision method \cite{KT_ex2}.}
\label{fig:tidal}
\end{figure}

Figure \ref{fig:tidal} plots the mass-shedding indicator $\chi$ as a
function of $\Omega m_0$ for the sequences of ${\bar M}^{\rm NS}_{\rm
B}=0.15$, and $Q=3$ and 5. For comparison, we also plot the results
obtained in the excision method. This figure shows that the curves
obtained in the moving-puncture approach agree approximately with those
in the excision approach. This indicates that the mass-shedding limit
would be determined with a small error even in the presence of spurious
eccentricity of order $\sim 0.1$. 

In the works performed in the excision approach \cite{KT_ex,KT_ex2}, the
ISCO is also determined by finding the extremum of the binding energy
(or the total angular momentum). They find that for ${\bar M}^{\rm
NS}_{\rm B}=0.15$ and $Q=5$, the binary reaches the ISCO before the
mass-shedding limit is reached, whereas for $Q=3$, the mass shedding
occurs before the binary reaches the ISCO. In the present
moving-puncture approach, we do not find any extremum even for
$Q=5$. This is primarily due to the fact that this approach is not
suitable for computing quasicircular states for very close orbits. 

We also compare our results of mass-shedding limit with that of the
dynamical simulations \cite{BHNS08}. Since it is difficult to determine
the onset of mass shedding in the dynamical simulations, we determine
the value of $\Omega m_0$ at the time when the matter in the NSs is
first swallowed by the BHs. For the case of $Q=3$, the matter is
swallowed by the BH at $\Omega m_0 \sim 0.09$, which is larger than the
mass-shedding limit obtained in this paper and in other works
\cite{KT_ex,KT_ex2}. However, two results are consistent because the
mass shedding occurs before the matter falls into the BH; $\Omega m_0
\sim 0.09$ shows the upper limit for the value of $\Omega m_0$ at the
mass shedding. On the other hand, for the case of $Q=5$, the matter is
first swallowed by the BH at $\Omega m_0 \sim 0.1$. This simply implies
that the binary reaches the ISCO at $\Omega m_0 \sim 0.1$.

\section{Summary} \label{sec:discussion}

We numerically derive new general relativistic quasiequilibrium states
of BH-NS binaries in the moving-puncture approach. Basic equations for
gravitational fields are solved in the mixture of conformal
thin-sandwich decomposition and conformal transverse-traceless
decomposition, and hydrostatic equations are solved under the assumption
of irrotational velocity field and employing the polytropic equation of
state.

In the moving-puncture approach, no definitive physical condition is
present for determining the center of mass of the system. We propose
three conditions for defining the center of mass and compare the
resulting quasiequilibrium states with the 3PN results and results
obtained in the excision approach.

Sequences of quasiequilibrium states are computed for the mass ratio $1
\le Q \le 5$, and for ${\bar M}^{\rm NS}_{\rm B} = 0.14$, 0.15, and 0.16
(each of which corresponds to the compactness ${\cal C}=0.1321$, 0.1452,
and 0.1600). For large orbital separation with $\Omega m_0\alt 0.02$,
the results in the moving-puncture approach agree with the results
obtained in the 3PN approximation and in the excision approach,
irrespective of the conditions for determining the center of mass and
irrespective of the mass ratio of the binary. Thus, such
quasiequilibrium states are likely to be (at least approximately)
quasicircular orbits for which the eccentricity is approximately
zero. By contrast, for small orbital separation with $\Omega m_0\agt
0.03$, the results in the moving-puncture approach with $\beta^\varphi$
and dipole conditions for determining the center of mass systematically
deviate from the 3PN results, in particular, for large mass ratio. The
angular momentum for the resulting quasiequilibrium is always smaller
than that in the 3PN results, and hence, the quasiequilibrium appears to
contain a nonzero eccentricity.  As an alternative method for
determining the center of mass, we propose a method in which it is
determined by requiring that the angular momentum for a give value of
the angular velocity should agree with that in the 3PN approximation.


To assess the circularity of the quasiequilibria computed in different
approaches, we performed numerical simulation for a chosen model;
$\bar M^{\rm NS}_{\rm B}=0.15$, $Q=3$ and 5, and $\Omega m_0\approx
0.033$. The numerical simulation was performed for two
quasiequilibrium states prepared in the $\beta^\varphi$ and 3PN-J
conditions. We find that the quasiequilibria computed in the
$\beta^{\varphi}$ condition have eccentricity of $\sim 0.05$--0.08.
During the simulation, the eccentricity reduces due to gravitational
radiation reaction, but in $\sim 4$--5 orbits, it does not reduce to
$\lesssim 0.01$ even at the onset of merger. By contrast, the
eccentricity of the quasiequilibrium state computed in the 3PN-J
condition is by a factor of $\sim 2$ smaller. For such case, the
eccentricity reduces approximately to $\sim 0.01$ in $\sim 4$--5
orbits, and the resulting eccentricity at the onset of merger appears
to be $\alt 0.01$. Therefore, with the quasiequilibrium prepared in
the 3PN-J condition, it is feasible to compute a realistic
gravitational waveform at least from the final a few inspiral orbits
to the merger phase (see an accompany paper \cite{BHNS08} for detailed
numerical results).


In this paper, we study BH-NS binaries in which the BH is not
spinning. If the BH has a substantial spin, the quasiequilibrium state
of BH-NS binaries will be modified by the spin-orbit interaction effect
(e.g. Ref.~\cite{3PN2}). We plan to study the effect of the BH spin in
the next work.

\begin{acknowledgments}
Numerical computation of quasiequilibrium states is performed using
the free library LORENE \cite{Lorene}.  We thank members in the Meudon
Relativity Group for developing LORENE.  This work was supported by
the Grant-in-Aid for Scientific Research (No. 19540263), by the
Grant-in-Aid for Scientific Research on Innovative Areas
(No. 20105004) of the Japanese Ministry of Education, Culture, Sports,
Science and Technology and by the Grand-in-Aid for JSPS. This work was
 also supported in part by NSF Grant No.PHY-0503366.
\end{acknowledgments}


\end{document}